\documentclass[11pt,a4paper]{article}
\usepackage[utf8]{inputenc}
\usepackage{amsmath}
\usepackage{amsfonts}
\usepackage{amssymb}
\usepackage{floatrow}
\usepackage{graphicx}
\usepackage{hyperref}
\usepackage{import}
\newcommand{\nn}{\nonumber}

\newtheorem{thm}{\indent{\sc Theorem}}

\newcommand{\benu}{\begin{enumerate}}
\newcommand{\eenu}{\end{enumerate}}

\newcommand{\ep}{\epsilon}

\newcommand{\beit}{\begin{itemize}}
\newcommand{\eeit}{\end{itemize}}

\newcommand{\be}{\begin{eqnarray}}
\newcommand{\beo}{\begin{eqnarray*}}

\newcommand{\ee}{\end{eqnarray}}
\newcommand{\eeo}{\end{eqnarray*}}
\newcommand{\e}{e^{\int_{0}^{t}\theta_u^2du}}

\title{A Mathematical Model of Foreign Capital Inflow}

\date{}

\author{Gopal K. Basak\thanks{Corresponding author: Stat-Math Unit, Indian 
Statistical Institute, Kolkata 700108, 
Email: gkb@isical.ac.in \ Research of this author is partially supported by the Institute's CPDA grant for 2014-2017.}, 
Pranab Kumar Das\thanks{Centre for Studies in Social Sciences, Calcutta, R1, 
B.P. Township, Kolkata 700094, India, Email: pkdas@cssscal.org
}, Allena Rohit\thanks{Emory University, Atlanta, USA, 
Email: rohit.allena@emory.edu}}
\begin{document}
{
\maketitle

\bigskip

\bigskip
\abstract{
The paper models foreign capital inflow from the developed 
to the developing countries in a stochastic dynamic 
programming (SDP) framework. Under some regularity 
conditions, the existence of the solutions to the SDP 
problem is proved and they are then obtained by numerical 
technique because of the non-linearity of the related 
functions. A number of comparative dynamic analyses explore 
the impact of parameters of the model on dynamic paths of 
capital inflow, interest rate in the international loan 
market and the exchange rate.
}
\bigskip

\noindent
\hspace{-30pt}

\medskip

\noindent
\hspace{-30pt}

\medskip

}

\section{Introduction}
The paper develops a simple model of foreign capital inflow in a multi period 
dynamic programming frame work. Typically capital flows from the developed to the developing countries owing 
to a higher rate of return in the latter because of scarcity of capital and 
thus a higher marginal product of capital  than in the developed countries 
which are capital rich with a lower marginal product of 
capital.\footnote{Marginal product of capital is directly related to interest rate in general and exactly equal in a competitive market framework. }\label{ft1aa} The foreign capital supplements the domestic 
resources of the developing countries for achieving a higher rate of capital 
formation and hence higher rate of economic growth. Generally developing 
countries are characterized by consistent current account deficit. This 
together with a higher interest rate creates a natural condition for flow of 
capital from the developed to the developing countries. The inflow to the 
borrowing country helps appreciate the domestic currency followed by rise in 
asset prices and local goods prices which via favourable fiscal conditions 
further encourages domestic credit creation. Repayment of loans in the next 
period has an adverse effect on exchange rate resulting into depreciation of 
exchange rate. This does not matter as long as inflow is sufficient. The 
crisis situation occurs particularly when the rate of capital inflow reaches 
a plateau or the global investors find a new country with even a higher rate 
of return then the new paradigm looks shop worn; a sudden stop a la Calvo 
(1998) to capital inflow or a reversal of current account balance leading to asset price contraction, decreased domestic investment and the economy adjusts 
backwards. In an extreme situation depreciation of domestic currency makes 
debt servicing difficult which can lead to a foreign exchange crisis or 
banking crisis or both (see Reinhart and Reinhart, 2008; Reinhart and Rogoff, 
2009; Wolf, 2008). \footnote{General empirical evidence can be found in Bordo 
and Meissner (2005); Bussi\`{e}re et al (2004) while country specific evidence can be found in Dominguez and Tesar (2005), Goldfajn and Minella (2005), Noland  
(2005). East Asian crisis of 1997 has been extensively discussed by Allen and 
Gale (2007), Krugman (1998), Gorton (2008), Rakshit (2002).}\label{ftac11}
The crisis may further get aggravated by bad macroeconomic 
management on fiscal and monetary fronts. The problem of debt servicing 
arises because capital inflow to developing countries are typically pegged in 
hard currency which has been termed as 'original sin' (Eichengreen and 
Hausman, 1999; Eichengreen, Hausman and Panizza, 2005). 

In a simple two period model the potential of financial crisis emanating from 
the depreciation of exchange rate at the time of repayment of foreign 
currency loans was modeled by Marjit, Das and Bardhan (2007). The argument is 
based on simple text book explanation of terms of trade effect based on trade 
theoretic argument as in Caves et al 1993; Helpman and Krugman, 1989. When all the producers sell in the international market 
at a given world price the combined supply reduces the world price though no 
individual seller or even a country can affect world price. The same argument 
applies in the context of capital inflow in a two period framework. It is argued that the terms of trade effect may dominate leading to a financial crisis. Basak, Das and Marjit (2012) extended the model by introducing repayment of 
previous period borrowing. It adds multi period dynamics in the 
structure for a better understanding of the possibility of crisis. Costinot et al. (2014) has argued for capital control on the basis of dynamic terms of trade in the same spirit as in this paper. The present paper is a formalization of the process of capital inflow in an infinite horizon set up with micro foundations of the choice problem of the agents - borrowers and lenders. 
The model investigates the effect of an increase in the perception of risk of default of international loan in the lending country, shift in the expectation of the future exchange rate or decreased productivity in the domestic sector in the borrowing country on capital inflow. The issue pertaining to the origin of the financial crisis in the process of inflow is also discussed. 

With this short introduction the paper proceeds as follows. Section 2 
formulates a model of capital inflow in a two country setup, a borrowing 
country and a lending country. From the individual optimization exercise we 
derive equilibria in the two markets, viz. international loan market and the 
foreign exchange market which are then solved for the interest rate 
in the international loan market and exchange rate by numerical method. Proofs of the existence of the solutions are given in the Appendices. In section 3, we undertake comparative dynamic exercises of changes in parameter values. The final section concludes.

\section{The Model}
We consider a two-country model with one borrowing (developing) and another 
lending (developed) country.  The two country framework is an abstraction 
from the reality where the two countries represent two country groups. 
The borrowing country group is capital scarce, thus has a higher marginal 
product of capital and hence a higher domestic interest rates whereas the 
lending country has lower domestic interest rates as this group is capital 
rich. \footnote{The borrowing country group resembles the emerging market 
economies among the developing countries with high growth potential such as 
China, India, Brazil, South Africa etc. These economies are attractive 
location for international capital compared to the least developed countries 
of Africa or Asia.}\label{ft1a} Actual borrowing-lending takes place through 
banks or financial intermediaries, there is no direct lending. A typical bank 
in the developing country borrows from both the domestic and international 
markets and lends to domestic production sector while a typical bank in the 
developed country borrows from domestic sector and lends to both the domestic 
and international borrowers. \footnote{In recent times a few emerging market 
economy firms have been borrowing directly in the international market, but 
major part of international borrowing takes place via banks or consortium of 
banks. External Commercial Borrowing, American Depository Receipts are two 
prominent examples of such direct borrowing by Indian firms in the 
international market.}\label{ft1b} The developing country banks have to repay the 
foreign debt along with the interest rate in the next period denominated in 
the lending country's currency, i.e., hard currency. International loan market is 
assumed to be competitive so that the international loan rate is given to 
both lending country and borrowing country banks. We do not consider the 
determination of domestic borrowing or lending rates of interest in the 
model. It is only the interest rate on foreign borrowing/ lending and the 
exchange rate that are determined in this model. Each bank in both the 
developed and the developing countries maximize discounted expected utility 
over infinite horizon, specifically an intertemporal mean-variance utility 
function, by choice of their portfolio of borrowing and lending. 

\subsection{Developed country}
Competitive banking in the developed country ensures that all banks are 
symmetric so that each one of them raise the same amount of deposit, assumed to be $K_t^D$ in each period. 
Number of banks is assumed to be $m^D$. We define, at time period $t$ 
$r_t^D$ = interest rate paid on deposits, 
$R_t^D$ = interest rate on domestic loans,
$R_t^*$ = interest rate on international loan,
$F_t$ = total funds at the disposal of the bank.
The bank lends $\mu_t$ fraction in the international market the rest in the 
international market with a risk of default $\epsilon_t$ for the foreign loan 
defined as 
\begin{equation*} 
\epsilon_t = \begin{cases} 1 & \mbox{with probability } p_t , \\ 
\epsilon_0 & \mbox{with probability } 1-p_t,  \ \epsilon_0 \in [0,1) .
\end{cases} 
\end{equation*}
Domestic loan has no risk of default. 
Total funds for lending $F_t$ is the sum of accumulated profit from the 
previous period and the amount of new deposits and is given by:
\begin{equation}
\label{eq3}F_{t}=F_{t-1}[(1 - \mu_{t-1})(1+R_{t-1}^{D})+(\mu_{t-1})(1+R_{t-
1}^{*})\epsilon_{t-1}]-(1+r_{t-1}^{D})K_{t-1}^{D}+K_t^D.\end{equation}
The expected profit of each bank in the current period is given by,
\begin{equation*}
E[\pi_{t}^{D}]=F_{t}[(1-\mu_t)(1+R_{t}^{D})+ 
\mu_{t}(1+R_{t}^{*})E(\epsilon_{t})]-(1+r_{t}^{D})K_{t}^{D},
\end{equation*}
and the variance by,
\begin{equation*}
V_{t}(\pi_{t}^{D})=F_{t}^{2} \mu_{t}^{2}(1+R_{t}^{*})^{2}V_{t}(\epsilon_{t}).
\end{equation*}
A typical bank chooses its portfolio to maximize its intertemporal expected 
profit adjusted for 
risk (measured by variance). The instantaneous utility (mean-variance utility 
function) of a bank in the developed country at time period t is 
\begin{equation}
\label{eq6}
\Omega_t^D=E_t(\pi_t^D)-\frac{\gamma}{2}V_{t}(\pi_t^D).
\end{equation}
The bank maximizes 
\begin{align*}
\begin{split}
[\Omega_{t}^{D}+B^{D}[E_{t}(\pi_{t+1}^{D})-
\frac{\gamma}{2}V_{t}(\pi_{t+1}^{D})]+{B^{D}}^{2}[E_{t}(\pi_{t+2}^{D})-
\frac{\gamma}{2}V_{t}(\pi_{t+2}^{D})]&+\\{B^{D}}^{3}[E_{t}(\pi_{t+3}^{D})-
\frac{\gamma}{2}V_{t}(\pi_{t+3}^{D})]+\ldots ,
\end{split}
\end{align*}
by choice of $\mu_t$ subject to the funds constraint, where $B^D$ is the 
discount factor assumed same for all time periods. 

  The above optimality problem is solved by Bellman optimization 
technique.\cite{Bellman} The problem of the bank can be stated in the Bellman 
framework as
\begin{equation}
\label{eq28}
V^{D}(F_{t})=max_{\mu_{t}}\{\Omega_{t}^{D}+B^{D}E_t(V^{D}(F_{t+1}))\},
\end{equation} subject to fund constraint \eqref{eq3} where $ V^{D}(F_{t})$ 
denotes the value function assumed to be  stationary. Since (i) the utility 
function is concave and the constraint
function is linear,  (ii) the utility function is time separable in 
contemporaneous 
control and state variables, and (iii) current decision affect current and 
future utilities
but not the past ones, Bellman principle can be applied here to solve the 
dynamic
programming problem. In particular, solution would be of the form of quadric 
function since the utility function is quadratic. \footnote{Value function 
can be approximated by Kalman filtering techniques,  but for simplicity a 
quadratic value function is assumed to solve the Bellman equation.}
\label{ftab1} 
So we let, 
\begin{equation}V^{D}(F_{t})=a+bF_{t}+cF_{t}^{2},
\end{equation} $a$, $b$ , $c$ are 
the coefficients of the value function to be determined.
It is to be noted that the value function needs to be concave to have a 
unique maximum which is satisfied if $c$ is negative.
\begin{thm}
\label{th1}
Under the linear quadratic framework, for the existence of unique positive 
solution $\mu_t$ for all $t$ 
it is necessary that $\frac{1}{(1+R_t^D)^2} < B^D < \min(1, \frac{(Z_t^2 + (1+R_t^*)^2 V_t(\ep_t))^2 }
{(1+R_t^D)^2 [(3/2) Z_t^2 (1+R_t^*)^2 V_t(\ep_t) + ((1+R_t^*)^2 V_t(\ep_t))^2]})
$
whenever $Z_t = (1+R_t^*)E_t(\epsilon_t) - (1+R_t^D) > 0$.
Further, for $\gamma \in (\gamma_0, \gamma_1)$, for some $0 < \gamma_0 < \gamma_1$,
the solution $\mu_t \in (0, 1)$ exists for all 
$t$ and is given by
\begin{eqnarray}
\label{eqth18}
\mu_t^* &=& \frac{Z_t[(1+B^{D}b)+2B^{D}cF_t(1+R_t^D)-2B^{D}cr_{tk}^D]}
{F_t[\gamma (1+R_t^*)^2V_t(\epsilon_t)-2B^{D}c(Z_t^2+(1+R_t^*)^2V_t(\epsilon_t))]}
\end{eqnarray}
with
\begin{equation*}
r_{tk}^D=(1+r_{t})K_t^D-K_{t+1}^D ,
\end{equation*}
where $b$ and $c$ are given in \eqref{eqb1} and \eqref{eqc1}.
\end{thm}

\bigskip
For the proof of this theorem please see Appendix A.
The intuition of the theorem obtains from the fact that
as $\mu$ increases from zero to positive values utility from out of lending 
increases but the marginal utility of lending falls with rising $\gamma$ 
(i.e., rising perception of risk) and at a high enough $\gamma$ the maxima of 
the utility function achieves at some $\mu<1$. Thus for a high enough 
$\gamma$, $\mu$ is bounded at unity from above.

\subsection{Developing country}
We also assume competitive banking in the developing country with total 
number of banks $m^U$ each of which is symmetric. Number of firms which 
borrow from each bank are normalized at unity.
A typical bank raises a fraction $\lambda_t$ from the international market at 
an interest rate $R_t^*$ and the rest from the domestic market at an interest 
rate $r_t^u$.  These funds are lent to domestic sector at an interest rates 
$R_t^u$ with a risk of default arising out of productivity shock denoted by 
$\eta_t$. A typical bank is assumed to hedge against currency fluctuations by 
buying a currency forward. Let us define 
$e_{t}$ = spot exchange rate at $t$, 
$e_t^f$ = forward market exchange rate,
$P_t^f$ = price of the forward contract at $t$. 
$G_t$ = funds available to each bank at $t$ (accumulated from past profits and new deposit raised) and is given by
\begin{equation*}
\label{constldc}
G_{t}=\eta_{t-1}(1+R_{t-1}^{U})G_{t-1}-K_{t-1}^{U}[(1-\lambda_{t-1})(1+r_{t-
1}^{U})+\lambda_{t-1}(1+R_{t-1}^{*})E_{t}(\frac{e_{t}}{e_{t-1}})]+K_{t}^{U}.
\end{equation*}
The profit of each bank in the developing country is thus given by
\begin{eqnarray}
\label{eq55}
\pi_t^U &=& \eta_{t}(1+R_{t}^{U})G_{t}-K_{t}^{U}[(1-
\lambda_{t})(1+r_{t}^{U})+\lambda_{t}(1+R_{t}^{*})(\frac{e_{t+1}}{e_{t}})] 
\nn\\ 
&& + (1+R_t^U)G_t(\frac{e_{t+1}}{e_t^f}-1)-P_t^f,
\end{eqnarray}
When foreign exchange markets are either informationally efficient or 
whenever risk premium is a non decreasing function of the level of 
transaction, forward contracts are priced in such a way that marginal benefit 
from the contract equals marginal cost of the contract. The last two terms in 
\eqref{eq55} vanish. Hence, the instantaneous mean variance utility function is given 
by
\begin{eqnarray*}
\Omega_t^U &=& E_t(\pi_t^U)-\frac{\beta V_t(\pi_t^U)}{2} \nn\\
&=& E_{t}(\eta_{t})(1+R_{t}^{U})G_{t}-K_{t}^{U}[(1-
\lambda_{t})(1+r_{t}^{U})+\lambda_{t}(1+R_{t}^{*})E_{t}(\frac{e_{t+1}}{e_{t}}
)] \nn\\
&& -\frac{\beta}{2}[(1+R_{t}^{U})^{2}G_{t}^{2}V_{t}(\eta_{t})
+(K_{t}^{U})^{2}\lambda_{t}^{2}(1+R_{t}^{*})^{2}V_{t}(\frac{e_{t+1}}{e_{t}})] .
\end{eqnarray*} 
The problem of the 
bank in the Bellman framework is given by
\begin{equation}
V^{U}(G_{t})=max_{\lambda_{t}}\{\Omega_{t}^{D}+B^{U}E_t(V^{U}(G_{t+1})\}.
\end{equation} subject to fund constraint \eqref{constldc}
where $V^{U}(G_{t})$is the value function assumed to be stationary and 
$B^{U}$ is the discount factor. 
As in the case of developed country the Bellman principle can be applied here 
on the quadratic value function to solve the dynamic programming problem. 
Thus,
\begin{equation}
\label{valueldc}
V^{U}(G_{t})=x+yG_{t}+zG_{t}^{2},
\end{equation}
with $x,y,z$ to be determined by equating coefficients. \footnote{Again, the 
parameter $z$ has to be negative for the same reason as explained earlier.}

\begin{thm}
\label{th2}
Under the linear quadratic framework 
for the existence of unique positive solution $\lambda_t$ for all $t$ 
it is necessary that 
$ \frac{1}{3 (1 + R_t^U)^2 V_t(\eta_t) +  (1 + R_t^U)^2 (E_t(\eta_t))^2} < B^U
< \frac{1}{(1 + R_t^U)^2 [ V_t(\eta_t) + (E_t(\eta_t))^2]} ,$
whenever $A_t=(1+r_t^U)-(1+R_t^*)E_t(\frac{e_{t+1}}{e_t}) > 0$.
Further, 
there exists $\beta$s neither very large nor very small
(in an interval, $(\beta_0, \beta_1)$, for some $0 < \beta_0 < \beta_1$) such
the solution $\lambda_t \in (0, 1)$ exists for all $t$ and is given by
\begin{eqnarray}
\label{eqth40}
\lambda_t^* &=& \frac{A_t[(1+B^{U}y)
+2B^{U}z(1+R_t^U)E_t(\eta_t)G_t-2B^{U}z r_{tk}^U]}
{K_t^{U}[(1+R_t^*)^2V_t(\frac{e_{t+1}}{e_t})
(\beta-2B^{U}z)-2B^{U}zA_t^2]}
\end{eqnarray}
with
\begin{equation*}
r_{tk}^U=K_t^U(1+r_t^U)- K_{t+1}^U
\end{equation*}
where $y$ and $z$ are given in \eqref{soly} and \eqref{solz}.
\end{thm}
The proof of this theorem is relegated to Appendix B.

It may be noted that for a small value of $\beta$ (implying low perception of 
risk) allows the developing country banks to borrow larger amount, possibly 
even $\lambda > 1$, as long as $R^*_t$ remains lower than the level given by 
$A_t>0$. On the other hand, if $\beta$ is too high then $\lambda$ can become zero or negative. Thus for a solution of $\lambda \in (0, 1)$ one must choose a $\beta \in (\beta_0, \beta_1)$, for some $0< \beta_0 < \beta_1$.
 
\subsection{Equilibrium: Existence and Characterization}  

There are two markets in this model, viz. international loan market and the 
foreign exchange market. Total supply of loans in the international 
market is the sum total of loans by the developed country banks and total 
demand for international loans is the sum total of demand by individual banks 
in the developing country. In equilibrium aggregate demand equals aggregate supply at each $t$ so that
\begin{equation}
\label{eq35}
m^D \mu_t^* F_t=\frac{m^U \lambda_t^* K_t^U}{e_t} .
\end{equation}
The R.H.S is aggregate demand and the L.H.S. is aggregate supply of loans. 
It may be noted that developing country banks cannot raise loans from the 
international
market at their home currency, it has to be in hard currency, such as US 
Dollar or EURO.
The equilibrium in the foreign exchange market is obtained by the equality of 
the demand for foreign exchange comprising of repayment of loans of previous 
period and the supply comprising of net exports plus current period loan. The current net exports is assumed to be a linear 
increasing function of current period exchange rate,
\begin{equation}
\label{eq100}
N_t=-N_0+N_1e_t
\end{equation}
where $N_1$ equals relative price of foreign goods (imports) vis-\`{a}-vis 
home goods (exports), with 
$N_1>0$ and both $N_0, N_1$ following uniform distribution in the intervals
$(N_{01},N_{02})$ and 
$(N_{11},N_{12})$ respectively and $N_0>0$. Stochastic coefficients of the net export function adds dynamics to the model. It is clear from \eqref{eq100} 
that net exports is actually rising function of price of foreign goods 
expressed in home currency relative to price of home goods. It may be noted 
that at zero exchange rate or a very low price of foreign goods relative to 
home goods there is a negative net exports, i.e. there is positive imports. 
The equilibrium 
in the foreign exchange market is given by 
\begin{equation}
\label{eq36}
-
N_{0}+N_{1}e_{t}+\frac{m^{U}\lambda_{t}^{*}K_{t}^{U}}{e_{t}}=\frac{m^{U}(1+R_
{t-1}^{*})\lambda_{t-1}^{*}K_{t-1}^{U}}{e_{t-1}^{*}} .
\end{equation}
The LHS gives the supply of foreign exchange while the RHS the demand for 
foreign exchange. 
  Equations \eqref{eq35}, \eqref{eq36} (with substitutions from 
\eqref{eqth18}, \eqref{eqth40}) simultaneously determine, $R_t^*$ and $e_t^*$ 
at each $t$
corresponding to realizations of random shocks in the respective periods.  
Accordingly we get the dynamics of capital inflow, interest rate and exchange 
rate. Existence of equilibrium solutions are given by the following theorem.
\begin{thm}
\label{th3}
Under the assumptions of Theorems \ref{th1} and \ref{th2}
there exist unique positive solutions to 
\eqref{eq35} and \eqref{eq36} provided
\begin{equation}
\label{condth3}
N_0^2 > 4 N_1 m^U K_t^U
\end{equation}
\end{thm}

\bigskip
For the proof of this theorem please see Appendix C.

The value function in each case is assumed to be stationary. 
 If $c<0$  and $z<0$ then the value function is concave which in turn 
ensures existence of unique maximum for each of the borrowing and lending
country banks respectively. It readily follows from \eqref{eqcns}that $c<0$ 
if and only if
$B^D (1+R_t^D)^2 > 1$ \ and \
$ B^D (1+R_t^D)^2  < \frac{(Z_t^2 + (1+R_t^*)^2 V_t(\ep_t))^2 }
{[(3/2) Z_t^2 (1+R_t^*)^2 V_t(\ep_t) + ((1+R_t^*)^2 V_t(\ep_t))^2} $
\ and \ $0<B^D<1$. 
So the condition is easily satisfied provided $B^D$ is small enough to make 
the product less than unity. Similarly $z<0$ if and only if 
$0<B^{U}(1+R_t^U)^2E_t(\eta^2)<1$ for the developing country banks. We need 
another condition on the interest rate on international loan for positive 
amount of foreign borrowing by developing country banks. 
The developed country bank lends only when the expected return from the loan 
is greater than the return from domestic deposits i.e.,
\begin{equation*}
(1+R_t^*)E_t(\epsilon_t)>(1+R_t^D) \ \ \mbox{is also a necessary 
condition for } \ \ \mu_t^*>0 .
\end{equation*}
The developing country bank borrows only when the  international interest 
rate adjusted for the exchange rate is less than the domestic interest rate 
i.e.,
\begin{equation*}
(1+R_t^*)E_t(\frac{e_{t+1}}{e_t})< (1+r_t^U) \ \ \mbox{is also a necessary 
condition for } \ \ \lambda_t^*>0 .
\end{equation*}
Combining these two conditions we have 
\begin{equation}
\label{condR} 
\frac{(1+R_t^D)}{E_t(\epsilon_t)}<(1+R_t^*)<\frac{(1+r_t^U)}{E_t(\frac{e_{t+1
}}{e_t})}.
\end{equation}
  In the next section, the endogenous variables, viz. $R_t^*$, $e_t $ and 
capital inflow are numerically solved for given initial values and 
parameters of the model and explain how the equilibria behave over time.

\subsection{Simulation}
 The two equations, \eqref{eq35} and \eqref{eq36}solve for equilibrium values 
of $R_t^*$ and $\mu_t$ for a given parametric configuration and  exogenous 
variables. We run the simulation for 30 time periods. It may be noted that 
the algorithm used for finding the equilibrium values of the endogenous 
variables employs a version of E-M Algorithm (Estimation Maximization employed in Maximum Likelihood Estimation in econometrics) in a 
dynamic setup.
Given below the initial values chosen for two countries and values 
for parameters of the model equilibrium values are calculated using the algorithm provided in Appendix D. We have set $B^D=0.91, B^U=0.8$, $G_0=10, K^U=20$ and $F_0=10, K^D=10$.   In the entire analysis we have assumed that the new deposits raised $K_t^U$ and $K_t^D$ are constant over time. 

\begin{table}[H] 
 \caption{Initial values of stochastic parameters} 
 
 \begin{tabular}{l l l l l l l l l l l l l} 
\hline
 
$E(\frac{e_{t+1}}{e_t})$&$E(\epsilon_t)$&$V(\epsilon_t)$&$V(\frac{e_{t+1}}{e_
t})$&$E(\eta_t)$&$V(\eta_t)$&$N_{01}$&$N_{02}$&$N_{11}$&$N_{12}$ \\
 \hline 
 0.92&0.94&0.09&0.25&0.85&0.09&1100&1200&15&18\\
 \hline
\end{tabular}
\end{table}

\begin{table}[H]
 \caption{Initial values of parameters and variables}
 \begin{tabular}{l l l l l l l l l l l l l l } 
\hline
 
$\gamma$&$\beta$&$m^D$&$m^U$&$K^D$&$K^U$&$R^D$&$r^D$&$R^U$&$r^U$&$R_0^*$&$e_0
$&$G_0$&$F_0$ \\
 \hline 
 4&1&10&100&10&20&0.05&0.04&0.2&0.15&0.14&75&10&10\\
 \hline
 \end{tabular}
\end{table}
The equilibrium $R_t^*$, $e_t^*$ and total capital inflow (international 
borrowing/ lending)
for 30 time period obtained via simulation are provided in Figs. \ref{fig2} 
through 
\ref{fig4} below. 

\bigskip
\begin{figure}[H]                                                      
\caption{Capital Inflow}
\includegraphics[width=0.9\textwidth]{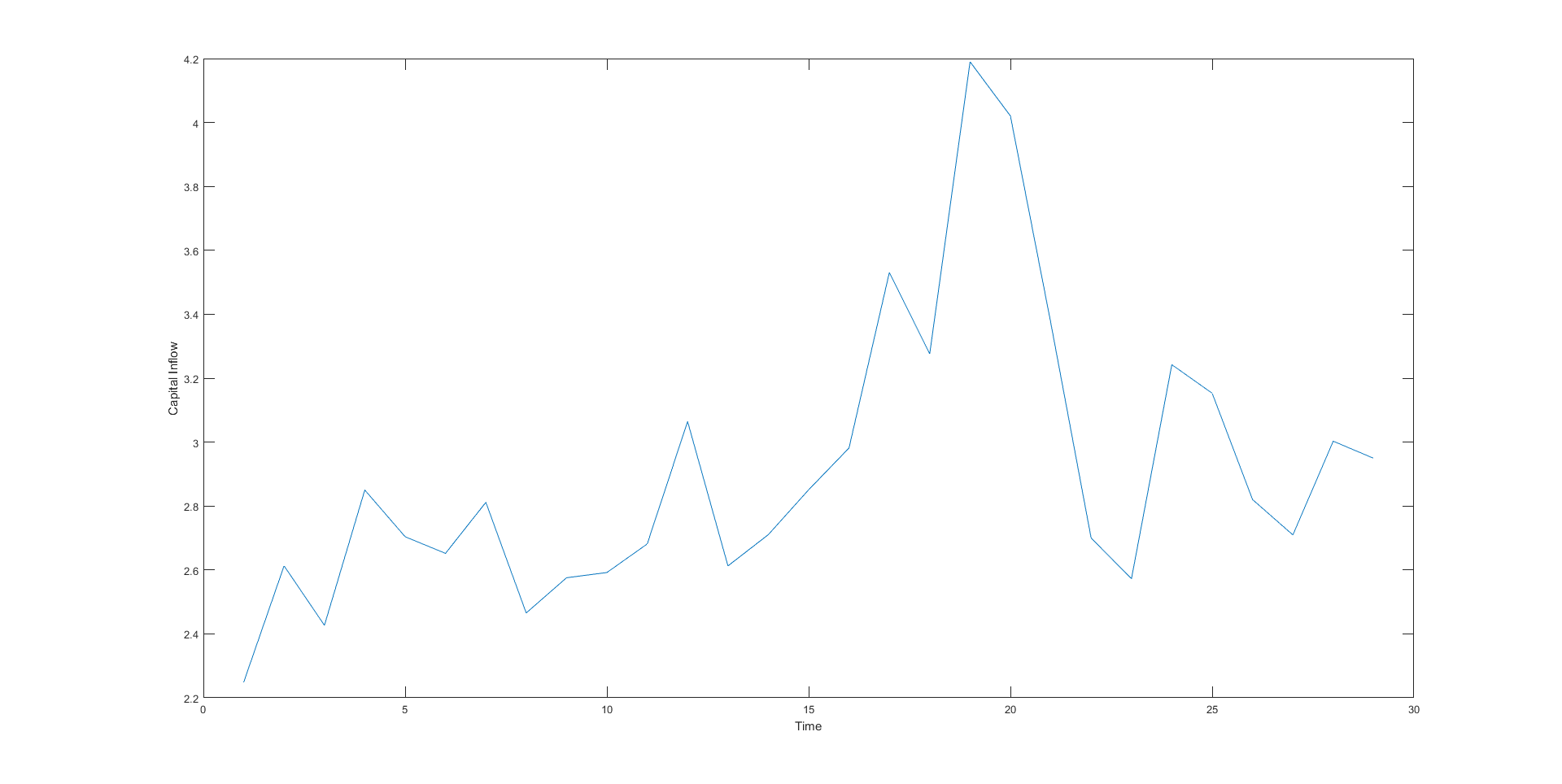}
\label{fig2}
\end{figure}

\begin{figure}[H]
\caption{$e_t$}
\includegraphics[width=0.9\textwidth]{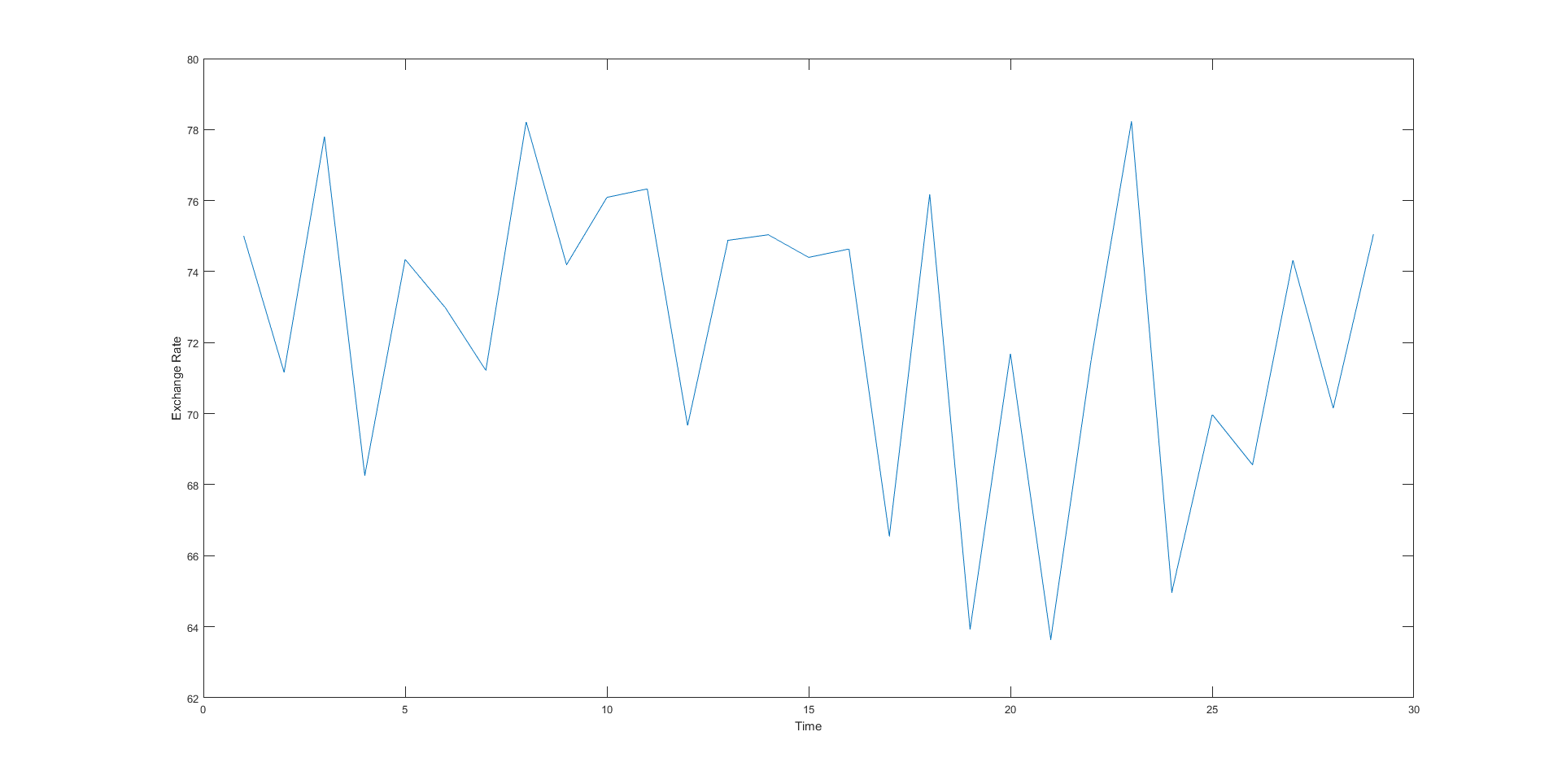}
\label{fig3}
\end{figure}

\begin{figure}[H]
\caption{$R_t^*$}
\includegraphics[width=0.9\textwidth]{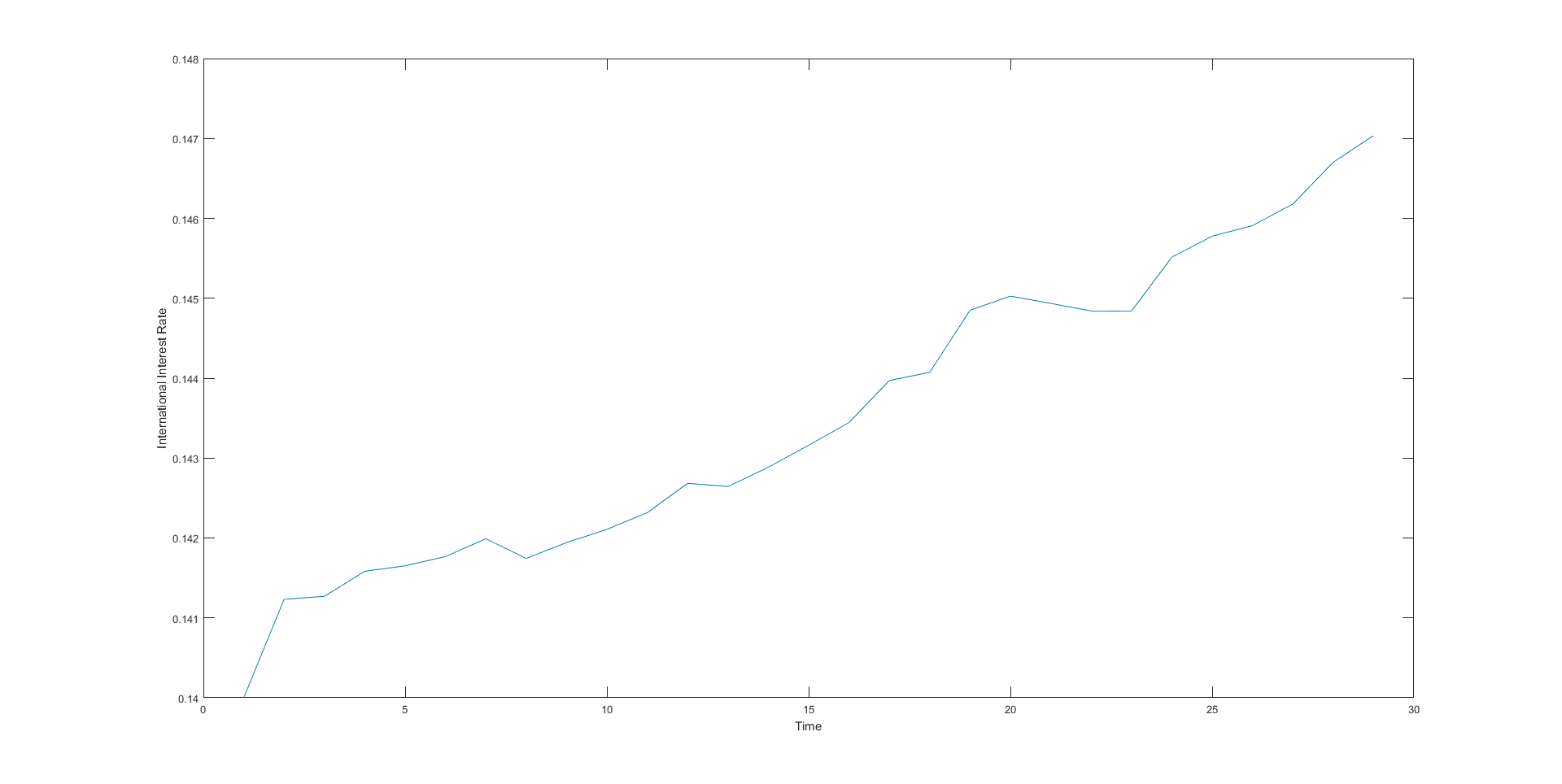}
\label{fig4}
\end{figure}
It is evident from the above figures that capital inflow and exchange rate 
follow  stationary pattern with a shift in the trajectory of the former between the period between 15 and 20. The international interest rate shows a rising 
trend in a small band.
This is because of the fact that a static expectation of next period exchange 
rate with a value less than unity implies banks in the developing country 
finds it optimal to borrow from the international market. The ever increasing 
demand for loans raises the interest rate over time. In the process the role of dynamic terms of trade operates via \eqref{eq36}. 
If there is an adverse supply shock in the foreign exchange market (due to, say a deficit in the balance of trade) then the second component on the L.H.S. of \eqref{eq36} has to fall to match the given value on the R.H.S. This interacts with the R.H.S. of \eqref{eq35} to achieve equilibrium in the foreign exchange market. The effect of current period exchange rate becomes more pronounced because the expectation of next period exchange rate is static - invariant to current information set about the future. \footnote{If the expectation formation is in tune with rational expectations hypothesis then current exchange rate also reflects expected exchange rate provided the foreign exchange market is efficient. In such a situation dynamic terms of trade effect may not be as strong as observed here.}
Eventually in the final equilibrium interest rate effect in each period 
dominates so that aggregate capital inflow to the developing country hovers 
around a stationary path. 
The fluctuation around the stationary path is governed by the stochastic 
shocks to net exports. 
In this model balance of trade has a dominant role in the determination of the intertemporal path of foreign capital inflow. As a matter of fact the trajectory of capital inflow is almost a mirror image of the trajectory of the exchange rate.

    In Table 3 below we provide the mean and variance of the endogenous 
variables. 
\begin{table}[H]
\caption{Mean and variance of the endogenous variables}
\begin{tabular}{ c c c c}
\hline 
Variable & Mean & SD & Coefficient of Variation\tabularnewline
\hline 
\hline 
$R_{t}^{*}$ & 0.1444 & 0.0015 & 0.0104\tabularnewline
\hline 
$e_{t}$ & 71.7915 & 4.4440 & 0.0619\tabularnewline
\hline 
Capital inflow & 3.0516 & 0.4532 & 0.1485\tabularnewline
\hline 
\end{tabular}
\end{table}

 The average exchange rate for the period of simulation turns out to be 
$71.7915$ units of domestic currency per unit of the foreign currency and the 
interest of the international market is $14.44\%$. The aggregate equilibrium 
capital inflow to the developing country is 3.0516 units. It may be noted that the mean and standard deviation of the 
endogenous variables reported in Table 3 are valid for the values of the parameters and 
initial values of the endogenous variables as in Tables 1 and 2. Any change 
in one or more of the parameter values generates different dynamic paths of 
the endogenous variables. Such exercises constitute comparative 
dynamics of the model. The new mean and standard deviation of the 
endogenous variables summarize the change in the series corresponding to new parameter values.   

\section{Comparative Dynamic Analysis}
We consider changes in some of the parameter values of interest for a better 
understanding of the economics of foreign capital inflow with emphasis on the 
primary focus of the model. It may be noted that for the comparative dynamic 
analysis we have retained the same realised values of shock to the net export 
function to make comparisons meaningful.
The set of baseline trajectory of the endogenous variables are plotted in 
black while the new set are plotted in red throughout this paper. 

\subsection{Change in risk perception of the developed country banks, 
$\gamma$} 
An increase in $\gamma$ 
implies that banks in the developed country perceive a higher risk for a 
given level of expected profit (see equation \eqref{eq6}). 
In this particular exercise $\gamma$ is raised from 4 to 15. The change corresponds to a dramatic shift in the risk perception and often  happens during the episodes of crisis. Then a typical bank in the developed country supplies lower amount of funds in the international market. As a result aggregate supply of foreign loans decreases resulting into increase in the international rate $R_t^*$. In the final equilibrium capital inflow to the developing country 
decreases and international rate increases for each $t$. 

\bigskip
\begin{figure}[H]                                                      
\caption{Effect of increased $\gamma$ on capital inflow (Bottom plot for 
higher $\gamma$)}
\includegraphics[width=0.9\textwidth]{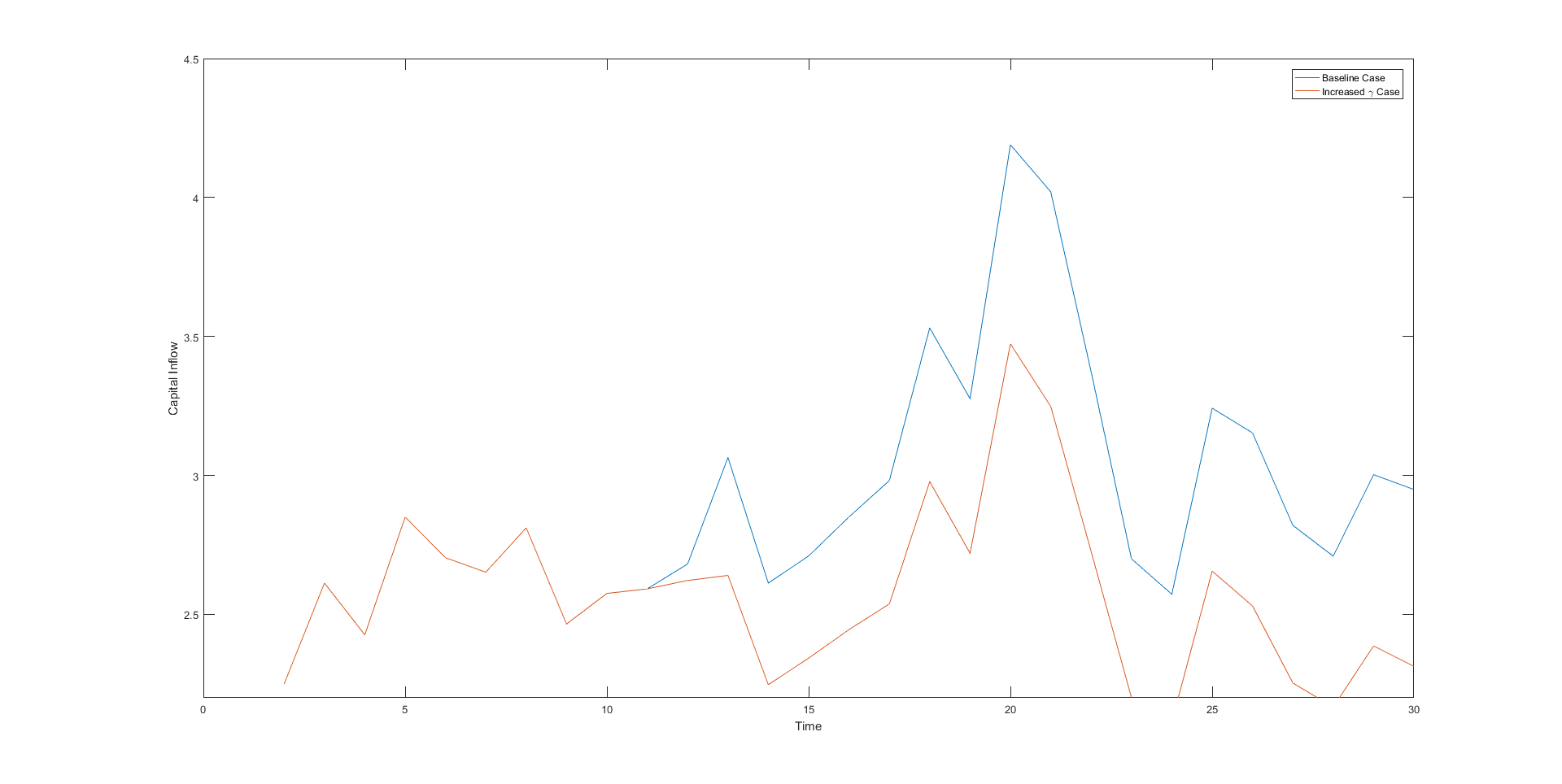}
\label{fig6}
\end{figure}

\begin{figure}[H]                                                      
\caption{Effect of increased $\gamma$ on 
$R^{*}_t$ 
(Top plot for higher $\gamma$)}
\includegraphics[width=0.9\textwidth]{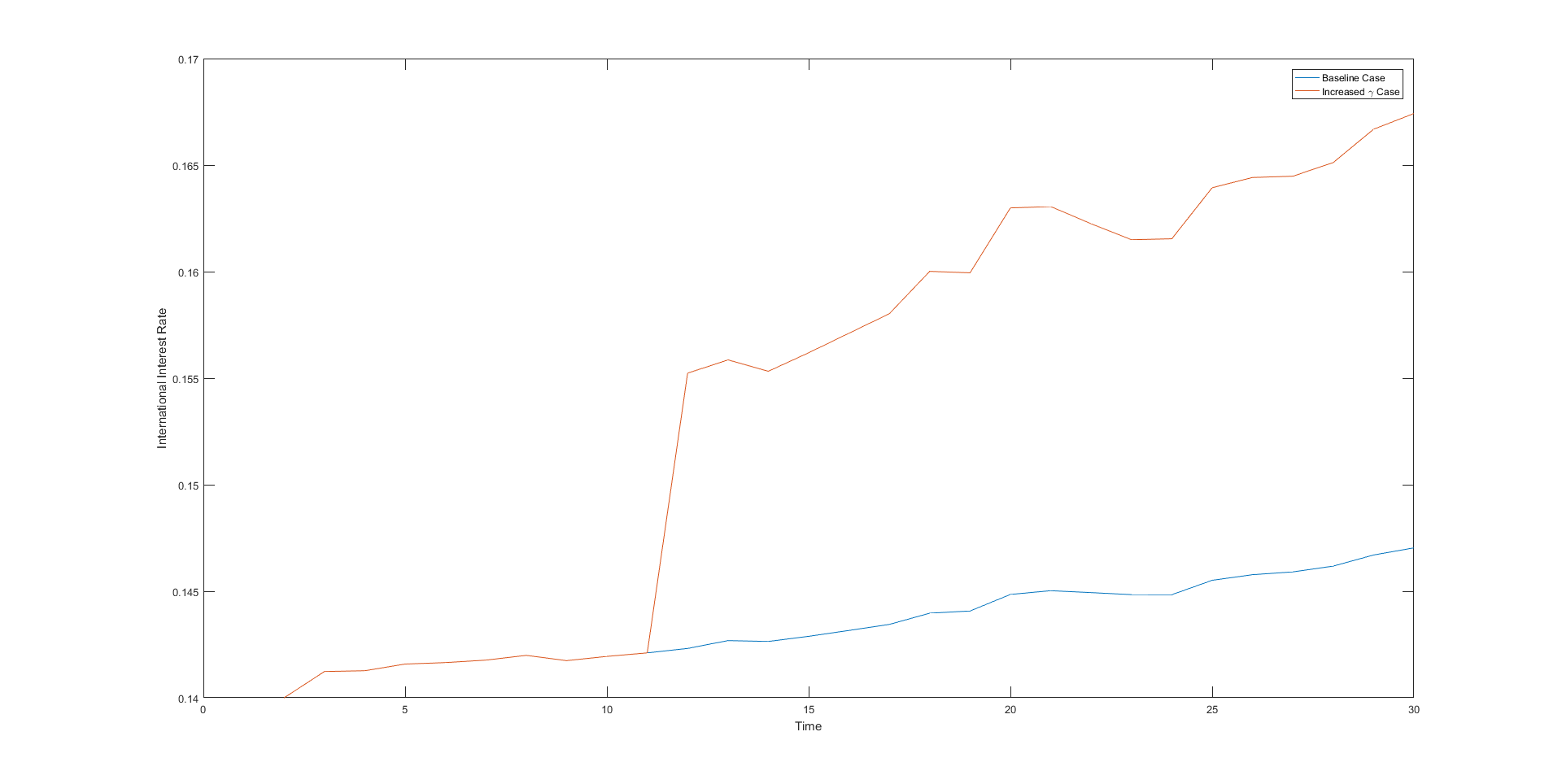}
\label{fig6a}
\end{figure}

\begin{figure}[H]                                                      
\caption{Effect of increased $\gamma$ on 
$e_t$
(difference with 
baseline)}
\includegraphics[width=0.9\textwidth]{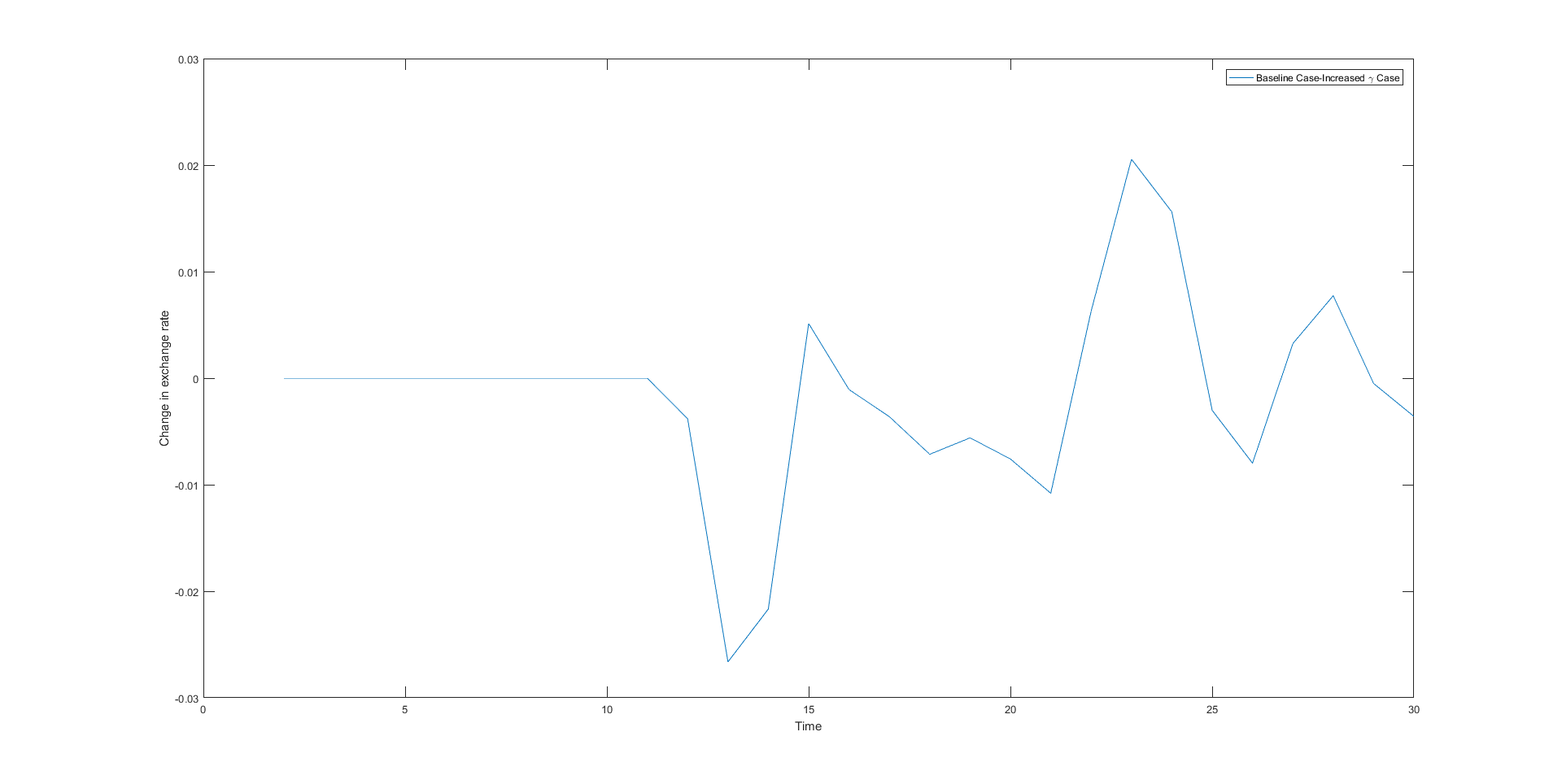}
\label{fig6b}
\end{figure}
Figs. 4 and 5 depict the new intertemporal trajectory (in red) of the capital 
inflow and the international interest rate along with the baseline (in 
black). The trajectory of capital inflow shows a downward shift from the 
original trajectory while the international interest rate shifts up. However, 
there is very little change in the exchange rate. To get a clear picture we 
plotted the difference of the new exchange rate from its baseline for each 
$t$ and plotted them in Fig. 6. The trajectory follows a stationary path 
around zero within a band from -0.027 to 0.02. There is no appreciable change in the 
exchange rate because the shock does not affect the net exports and so the 
balance of trade which as we argued earlier has a dominating role in the 
determination of the equilibrium trajectories.

\begin{table}[H]
\caption{Mean and variance of variables-increased $\gamma$} 
\begin{tabular}{c c c c }
\hline 
Variable & Mean & SD & Coefficient of Variation\tabularnewline
\hline 
\hline 
$R_{t}^{*}$ & 0.1602 & 0.0057 &  0.0356\tabularnewline
\hline 
$e_{t}$ &  71.7937 &  4.4425 & 0.0619 \tabularnewline
\hline 
capital inflow &  2.5597  & 0.3543  & 0.1384 \tabularnewline
\hline 
\end{tabular} 
\end{table} 

The averages for the new values of the endogenous variables also confirm 
this. There is a decrease in the average foreign capital inflow from $3.0516$ 
to $2.5597$ which is expected because of the increase in borrowing rate 
$R_t^*$. However, the percentage fall in foreign capital inflow 16.12\% is 
more than  the rise in $R_t^*$ 10.94\%. The increase 
in risk perception, $\gamma$ is  275\% from its initial value. The rate of 
change in the average capital inflow with respect to $\gamma$ is -0.0447 
while the rate of change of $R_t^*$ is 0.00014. 
The elasticity of capital inflow and interest rate with respect to $\gamma$ are -0.0379 and 0.0014 (evaluated at sample averages). Clearly the sensitivity of capital inflow to risk perception is much higher than the sensitivity of interest rate.

\subsection{Effect of a rise in $E_t(\frac{e_{t+1}}{e_t})$: Expected 
depreciation of exchange rate}
When banks in the developing country (borrowers in the developing country in 
general) expect a depreciation of exchange rate in the future, meaning an 
increase in the expected exchange rate vis-\`{a}-vis current rate, 
$E_t(\frac{e_{t+1}}{e_t})$, cost of borrowing (in terms of foreign 
currency) increases.  This change in expectation leads to a decrease in the 
demand for foreign borrowing which in turn reduces $R_t^*$ in the final 
equilibrium in each period. $E_t(\frac{e_{t+1}}{e_t})$ is raised from 0.92 to 
0.98 (6.52\% rise), with no change in the other parameter values. Comparing 
the new intertemporal trajectory with the baseline trajectory we find that 
the capital inflow attains a low level trajectory though the nature of the time 
path remains similar to the baseline trajectory.

\bigskip
\begin{figure}[H]                                                      
\caption{Effect of depreciation of 
$E(\frac{e_{t+1}}{e_t})$
on capital inflow 
(Bottom plot for higher $E(\frac{e_{t+1}}{e_t})$)}
\includegraphics[width=0.9\textwidth]{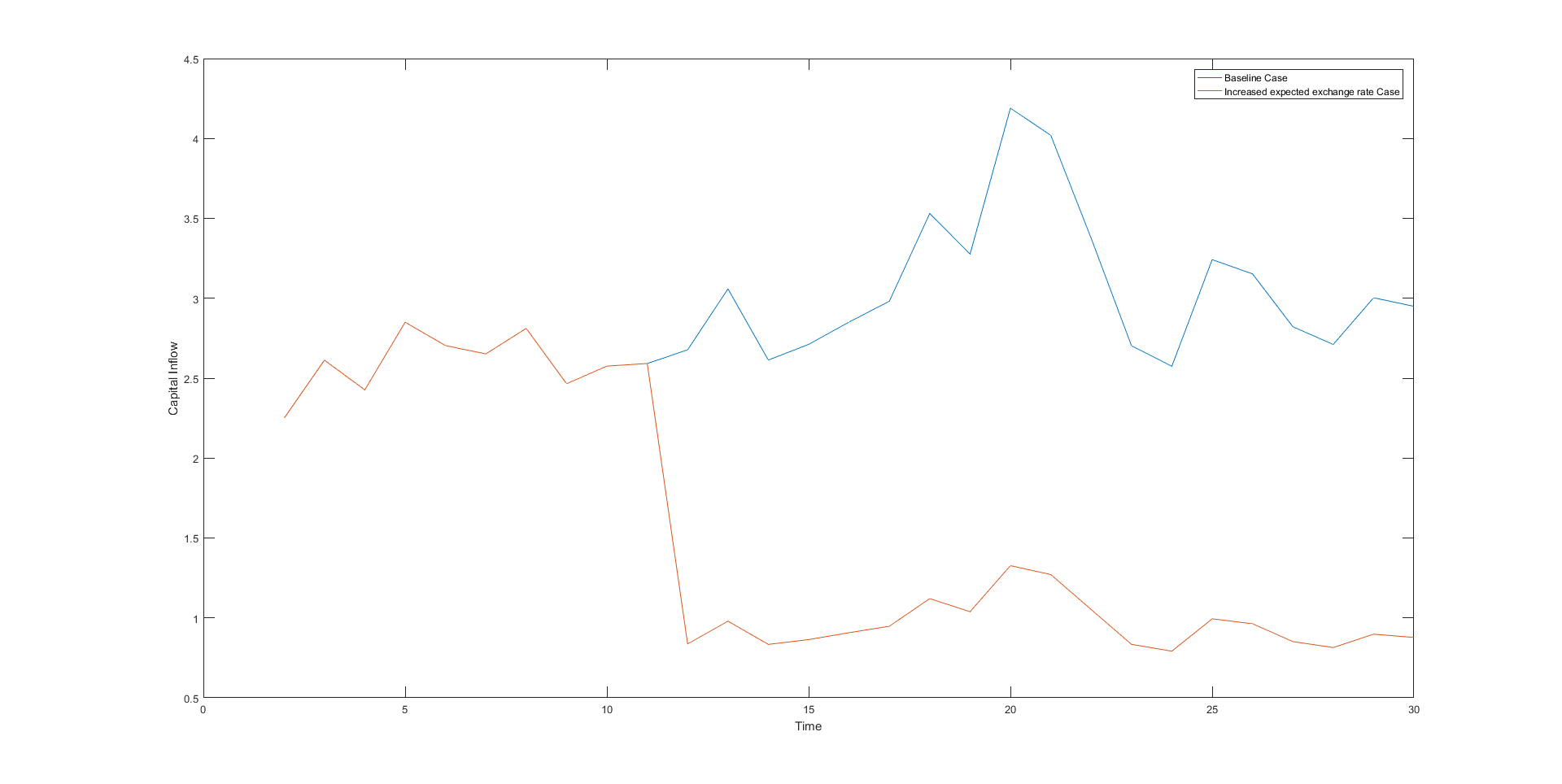}
\label{fig6c}
\end{figure}
\begin{figure}[H]                                                      
\caption{Effect of depreciation of 
$E(\frac{e_{t+1}}{e_t})$
on 
$e_t$
(difference from baseline)}
\includegraphics[width=0.9\textwidth]{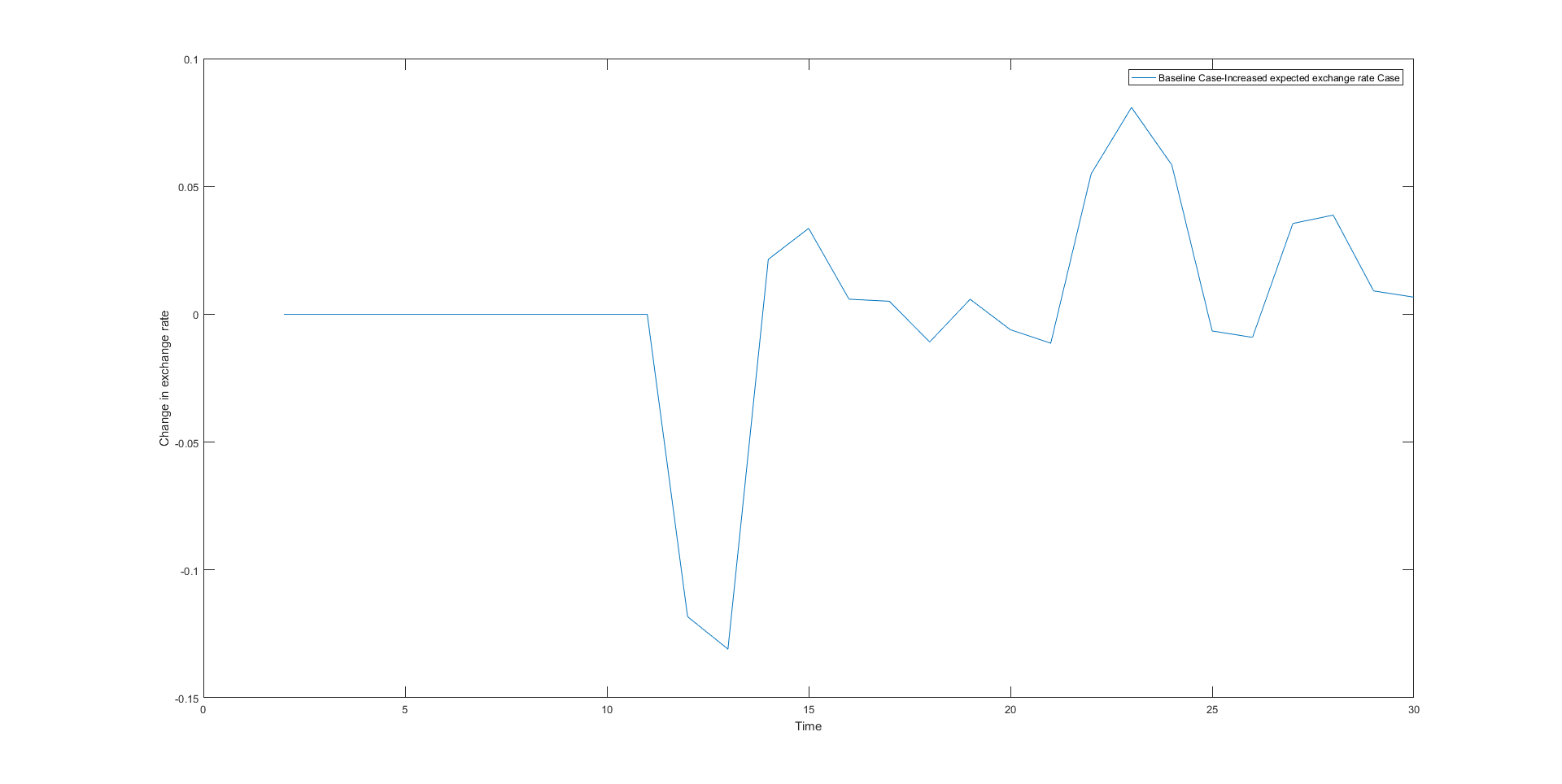}
\label{fig7}
\end{figure}
\begin{figure}[H]                                                      
\caption{Effect of depreciation of 
$E(\frac{e_{t+1}}{e_t})$
on 
$R^{*}_t$
 (Bottom plot for higher $E(\frac{e_{t+1}}{e_t})$)}
\includegraphics[width=0.9\textwidth]{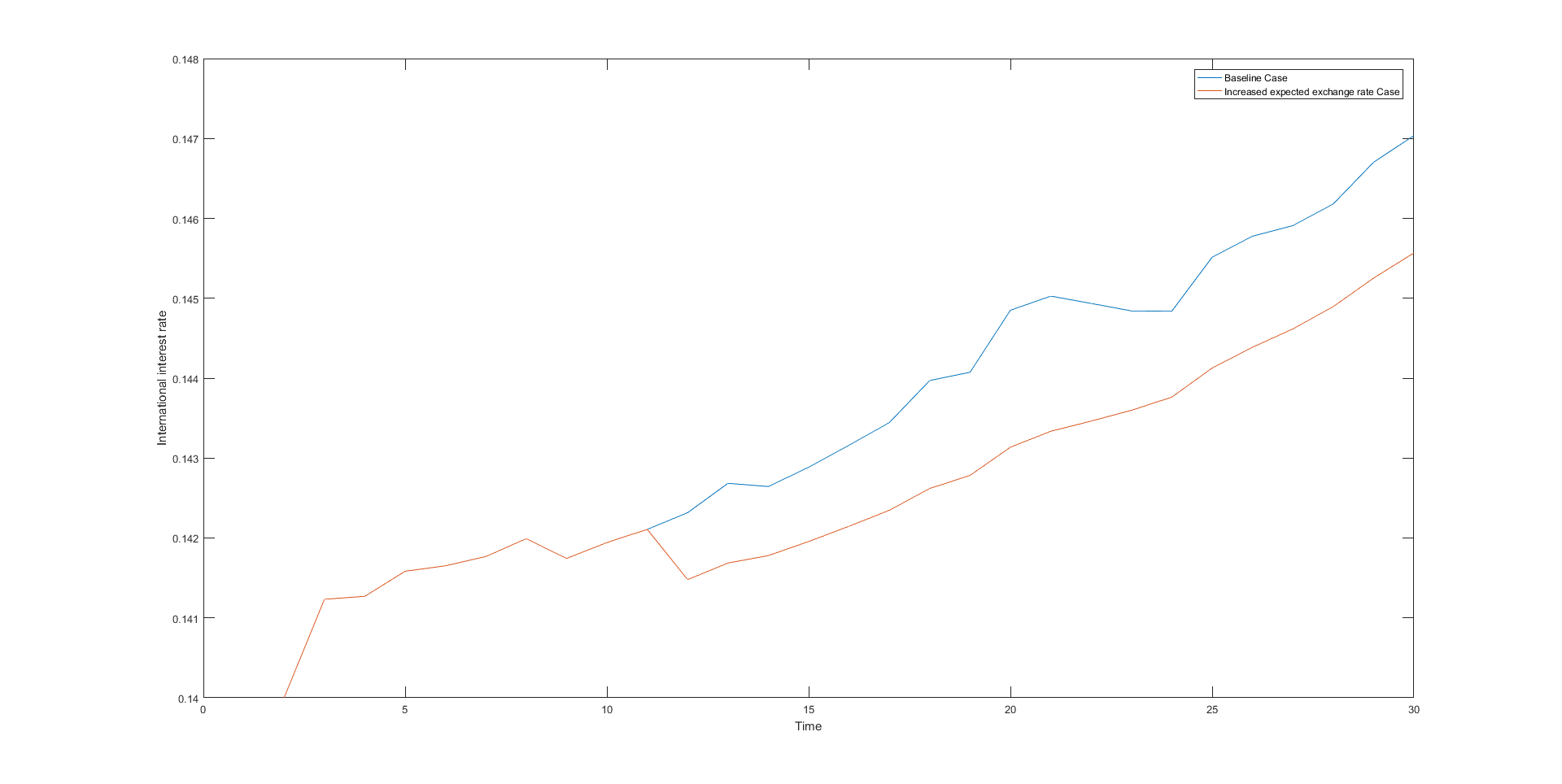}
\label{fig8}
\end{figure}

\begin{table}[H]
\caption{Mean and variance of variables - Increased $E(\frac{e_{t+1}}{e_t})$}
\begin{tabular}{c c c c }
\hline
 Variable & Mean & SD & Coefficient of Variation\tabularnewline
\hline 
\hline 
$R_{t}^{*}$ & 0.1433 & 0.0013 &  0.0087\tabularnewline
\hline 
$e_{t}$ & 71.7883 &  4.4439 & 0.0619 \tabularnewline
\hline 
capital inflow &  1.0399  & 0.3934  & 0.3783 \tabularnewline
\hline 
\end{tabular} 
\end{table}

Comparing the summary statistics reported in Table 5 with the baseline 
summary statistics reveals that capital inflow decreases by 65.92\% and 
$R_t^*$ decreases by 0.76\% on an average. The rate of change of capital inflow with respect to expected exchange rate is -33.53 which is much higher than the rate of change of interest rate (-0.0183). The elasticities of capital inflow and interest rate with respect to expected exchange rate are found to be -15.57 and -0.121. The current exchange rate does not change 
to any appreciable level. Difference of the current exchange rate from the baseline plotted in Fig. 8 shows that the range of variation is not significant (from $-0.14$ to $0.07$). The behaviour of the 
endogenous variables as depicted in the figures are also confirm from the 
respective averages over the periods provided in Table 5. 

\subsection{Decrease in the productivity of the domestic sector of the 
developing country, $E_t(\eta)$}
When there is a decrease in the productivity of the domestic sector of the 
developing country, $E_t(\eta_t)$, profit in the production sector falls 
which in turn leads to decreased demand for foreign loans by the developing 
countries. We conduct this exercise by reducing $E_t(\eta_t)$ reduced from 0.85 to 0.70 (roughly 17.64\% decrease). 
Given below the plots of endogenous variables with the new $E_t(\eta_t)$.

\bigskip
\begin{figure}[H]
\caption{Effect of a lower $E_t(\eta)$ on Capital Inflow (Bottom plot for 
lower $E_t(\eta)$)}
\includegraphics[width=0.9\textwidth]{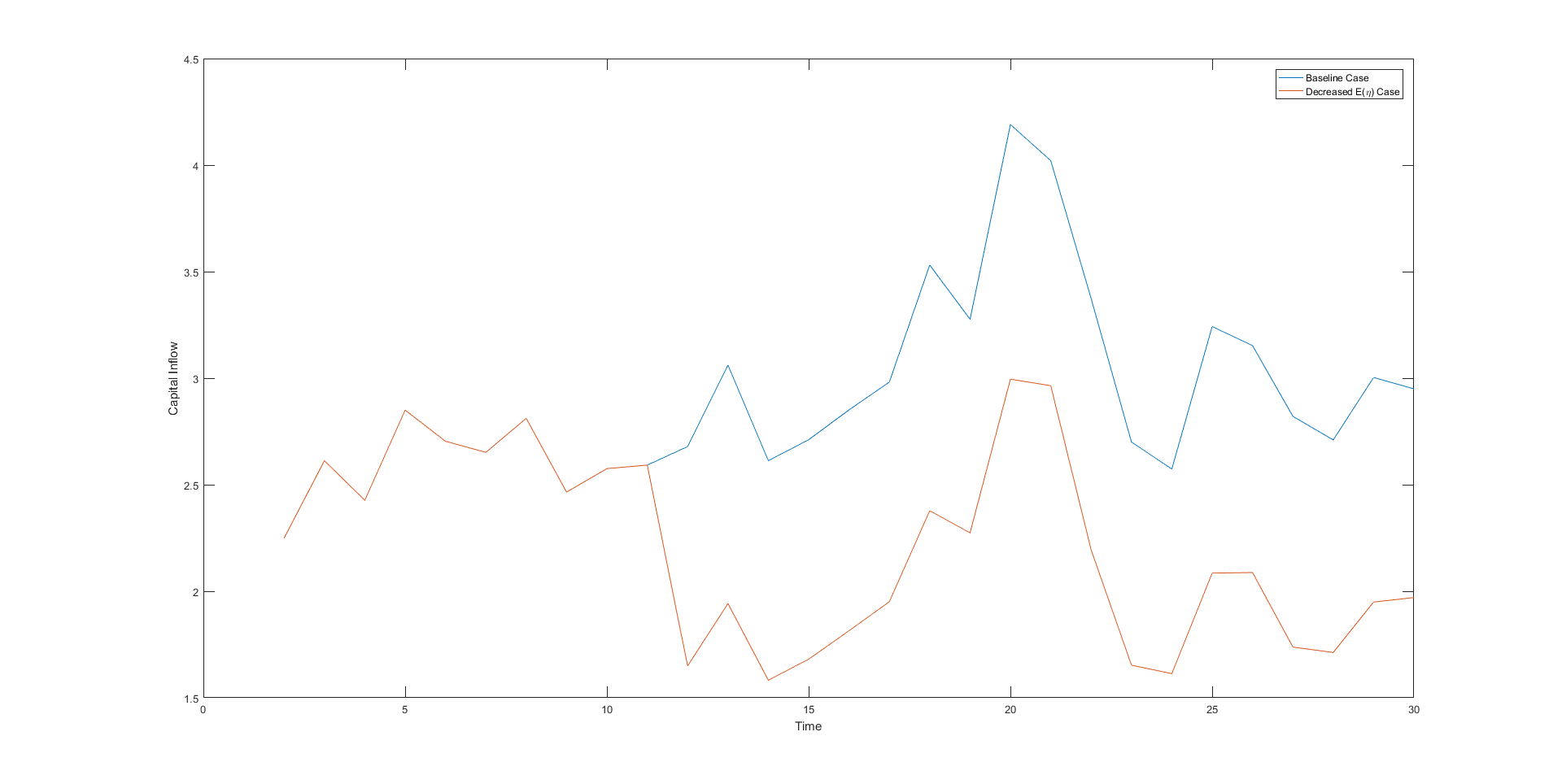}
\label{fig10}
\end{figure}

\begin{figure}[H]
\caption{Effect of a lower $E_t(\eta)$ on $R_t^*$ (Bottom plot for lower 
$E_t(\eta)$)}
\includegraphics[width=0.9\textwidth]{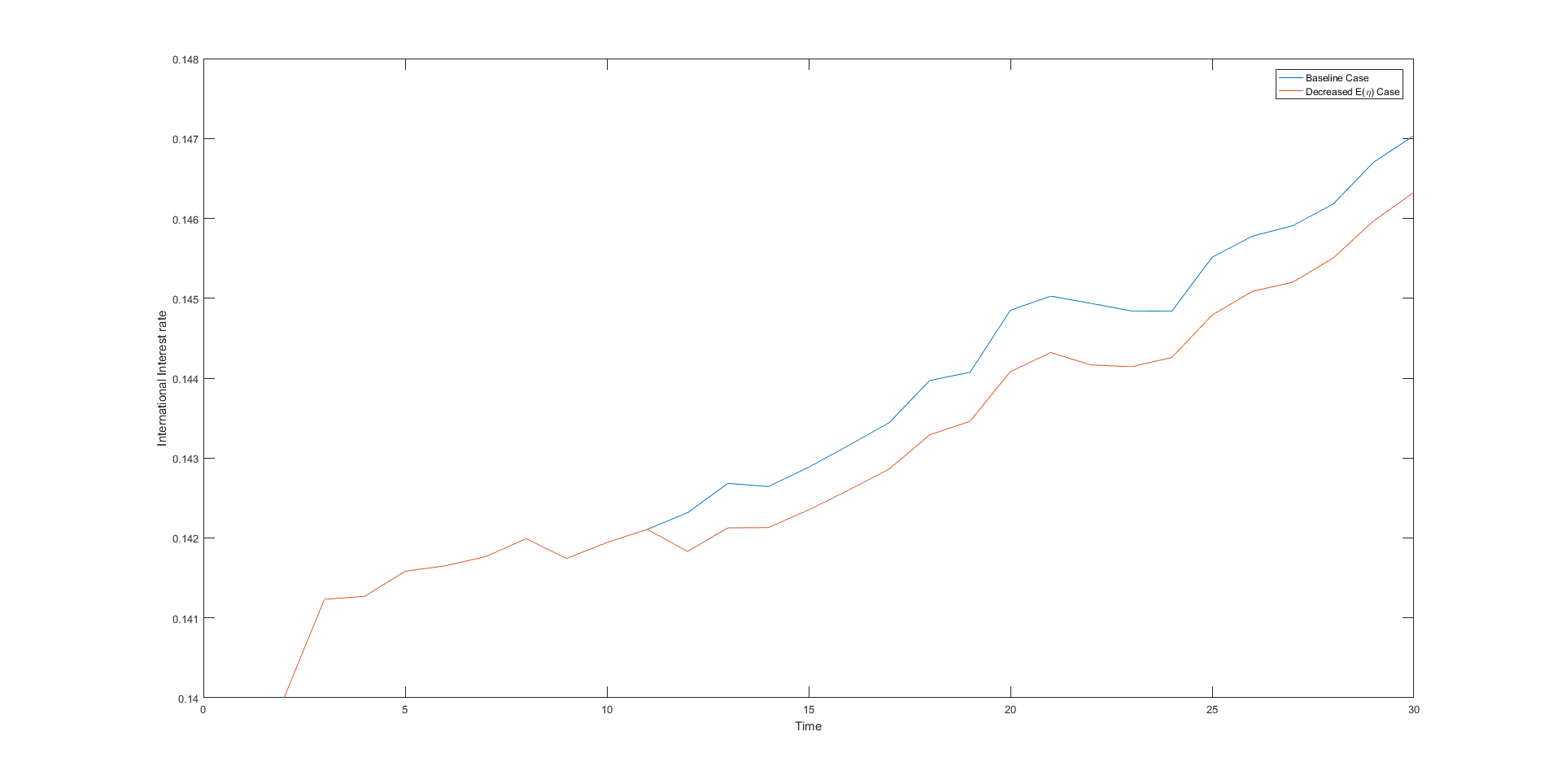}
\label{fig11}
\end{figure}

\begin{figure}[H]
\caption{Effect of a lower $E_t(\eta)$ on $e_t$ (in difference)}
\includegraphics[width=0.9\textwidth]{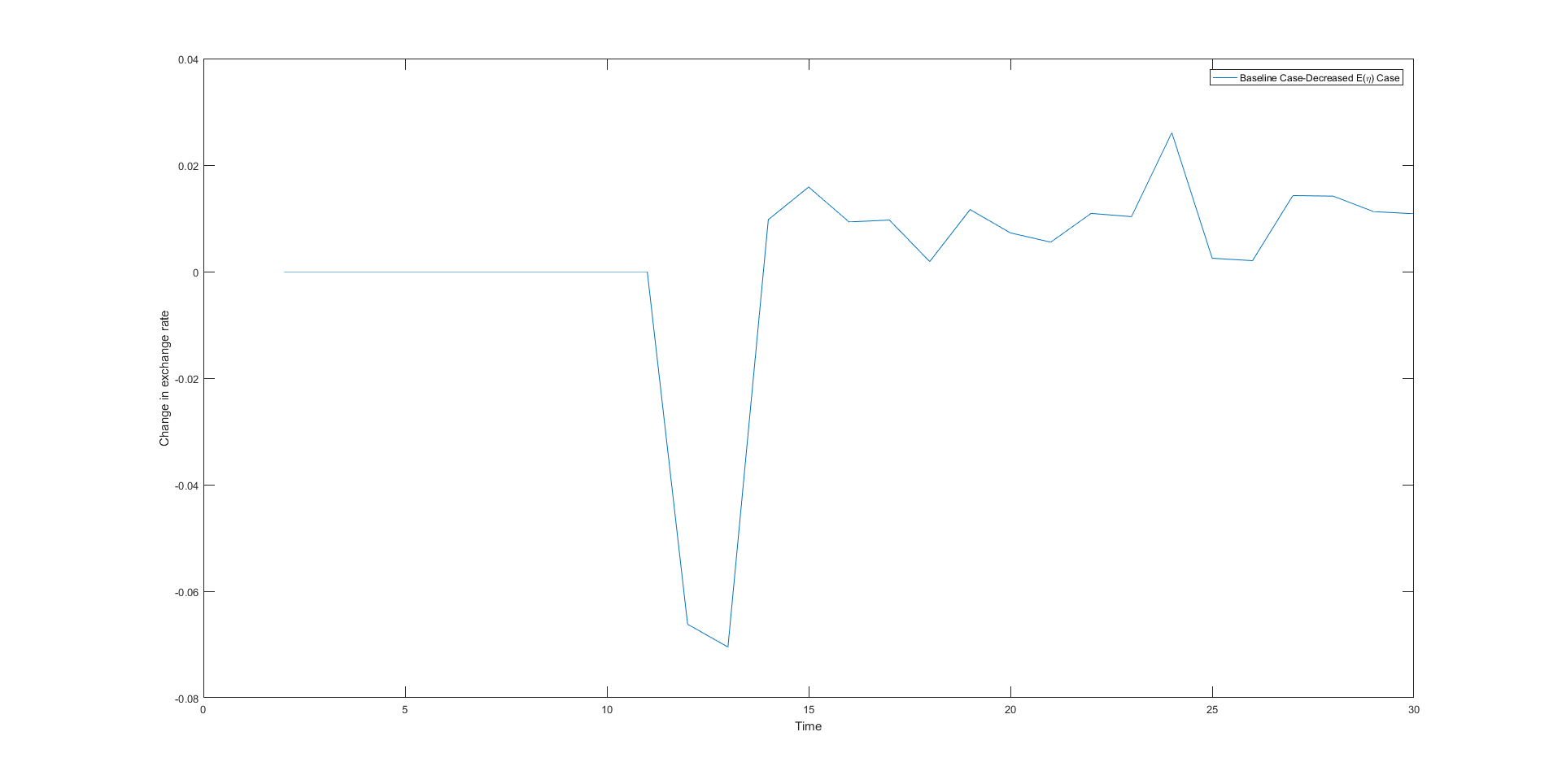}
\label{fig11a}
\end{figure}
From the plot it is seen that the whole trajectory of $R_t^*$ as well as the 
trajectory of the capital inflow shift below. Comparing the averages with that of the baseline (see Table 6) it is evident that there is no appreciable change in $e_{t}$ while there is fall in both the capital inflow and $R_{t}^{*}$ by 6.73\% and 0.004\% respectively. Thus the fall  capital inflow is much higher and the fall in $R_{t}^{*}$ is a slightly lower compared to the decrease in productivity, $E(\eta_t)$. The rate of change of capital inflow with respect to $E(\eta_t)$ is 6.7 while that for $R_{t}^{*}$ is 0.004 and the corresponding elasticities are 2.05 and 0.022 respectively.
\begin{table}[H]
\caption{Mean and variance of variables-shock to $E_t(\eta)$} 
\begin{tabular}{c c c c }
\hline 
Variable & Mean & SD & Coefficient of Variation\tabularnewline 
\hline 
\hline 
$R_{t}^{*}$ 
& 0.1438 & 0.0014 &  0.0096\tabularnewline
\hline 
$e_{t}$ 
&  71.7896 &  4.4437 & 0.0619 \tabularnewline
\hline 
capital inflow 
&  2.0415  & 0.4212  & 0.2063 \tabularnewline
\hline 
\end{tabular} 
\end{table}

\bigskip
An increase in the production cost in the developing country has similar 
effect as the decrease in the productivity. Such increases can occur due to 
increase in the price of inputs, wages, interest rate or price of raw 
materials. A very common cause of increased production cost arises due to 
increase in the prices of imported inputs, particularly crude petroleum and generates balance of payments crisis. 
The comparative dynamic analyses considered above show that the international 
interest rate is less sensitive to changes in the parameter values than 
capital inflow. This is true whether the parametric 
changes occur in the developed country or in the developing country.  A comparison of the elasticities of capital inflow and rate of interest in the international loan market with respect to the parameter values reveals that responsiveness with respect to the expected exchange rate in the 
next period far exceeds the responsiveness with respect to the risk perception of loans of the developed country banks or productivity of the domestic sector of the developing country (borrowing firms in the developing country). However, the behaviour of the current exchange rate is 
governed by the behaviour of the net exports. It is seldom affected by the 
change in the expected exchange rate. This is because of the fact that the 
nature of expectation of the exchange rate is static, expectation of future 
exchange rate is invariant with respect to behaviour of the current exchange 
rate. 

The above analyses can be useful for an analysis of the potential of 
financial crisis in this model. All the above three cases have the potential 
for generating financial crisis when foreign capital inflow drops to a very low level; in the extreme case it can hit the horizontal axis as in Figs. 4 and 10 leading to what can be called sudden stop a la Calvo (1998). As a matter of fact in the case of a sudden stop it is very difficult for the borrowing country to come back to its original position. However, the crisis, in this model does not occur because of imperfection in the loan market resulting into binding finance constraints for firms which becomes profound with goods price rigidity as in Aghion et al(2000). It is not the fallout of classic problem of moral hazard or adverse selection (Krugman, 1998). Nor it is (mis)information issue that has been ascribed very important role in the sense of herd behaviour discussed by Kindleberger (1978) and formally modelled by Banerjee (1992).

We have shown that a favourable foreign capital inflow can turn 
into a crisis even when macroeconomic fundamentals are good. In view of the 
such vulnerability of unrestricted capital flow to financial crisis eminent 
economists such as Bhagwati (1998), Ostroy et al (2010), Rodrick (1998) have 
suggested policy to tax capital out flow, such as Tobin type tax in general 
even when macroeconomic  fundamentals show good performance and Sen (2006) in 
particular has argued against full capital account convertibility for the Indian 
economy. 

\section{Conclusion}
The study addresses the economics of foreign capital inflow based on trade 
theoretic explanation operating via terms of trade effect in a dynamic 
context. The model is set up and solved in a two country group framework -  
developing or borrowing group and developed or lending
group. Each individual agent in both the country groups chooses her respective loan portfolio for each period by optimization of an intertemporal objective function of mean-variance variety. The equilibrium in the loan market together with the foreign exchange market solves the endogenous variables of the model using numerical methods. Net exports which constitutes the supply side of the 
foreign exchange market dominates intertemporal trajectory of endogenous 
variables. Simulation exercises are then conducted to derive   comparative 
dynamic results to assess the impact of shocks to parameters of the model on 
the intertemporal trajectory of the endogenous variables, viz. international 
interest rate, foreign capital inflow and exchange rate.

 Given the structure of the model a change in the risk perception in the 
developed country reduces capital inflow to the developing country group as a 
whole. An increase in the expected exchange rate vis-\`{a}-vis the current 
exchange rate raises the effective cost of foreign loans leading to a 
decrease in demand which in the final equilibrium also reduces the 
international interest rate.  A fall in the productivity in the domestic 
sector of the developing country leads to a fall in foreign capital inflow 
and a fall in the international interest rate. However, the responsiveness with respect to expected exchange rate vis-à-vis current exchange rate is much more compared to the other comparative dynamic analyses considered here.

The paper can be extended in several directions. A restrictive assumption of 
the model is static expectation of exchange rate. An extension with 
endogenous determination of expected exchange rate is capable to add richer 
dynamics to the model. Another interesting extension could be the endogeneity 
of risk of default that is dependent on the outstanding borrowing of the 
developing countries. Finally, a multi country extension within the borrowing 
country group can be undertaken to show the contagion effect when a shock 
originating in one of the borrowing countries spreads to the other members of 
the group.
 
\section{Appendix A}
\label{app1}
\setcounter{equation}{0}

{\bf Proof of Theorem \ref{th1}:} \ 
From \eqref{eq28}  
\begin{equation*}
a+bF_{t}+cF_{t}^{2}
=max_{\mu_{t}}\{\Omega_{t}^{D}+B^{D}E_t(V^D(F_{t+1})\},
\end{equation*}
where 
\begin{equation*}
\Omega_{t}^{D}=F_{t} (1-\mu_{t})(1+R_{t}^{D})+F_{t} \mu_t 
(1+R_{t}^{*})E_{t}(\epsilon_{t})-(\gamma/2)F_{t}^{2} 
\mu_t^{2}(1+R_{t}^{*)^{2}}V_{t}(\epsilon_{t})
\end{equation*}
 and
\begin{equation*}
E_{t}(F_{t+1})=F_{t}[(1-\mu_{t}) (1+R_{t}^{D})+ \mu_t 
(1+R_{t}^{*})E_t(\epsilon_{t})]-(1+r_{t}^{D})K_{t}^{D}+K_{t+1}^{D} .
\end{equation*}
To solve this, denote  
$M_{t}^{D}=\Omega_{t}^{D}+B^{D}[a+bE_{t}(F_{t+1})+cE_{t}(F_{t+1}^{2})]$, 
then the value function takes the form,
\begin{equation*}
V^{D}(F_{t})
=\mbox{ma\ensuremath{x_{\mu_{t}}}\{\ensuremath{M_{t}^{D}\}}}.
\end{equation*}
First order condition to the above equation yields 
\begin{equation}
\frac{\partial M_{t}^{D}}{\partial\mu_{t}^*}|\mu_{t}^*=0 .
\end{equation}
Now, 
\begin{equation*}
\frac{\partial \Omega_{t}^{D}}{\partial\mu_{t}^*}=
-F_{t}(1+R_{t}^{D})+F_{t}(1+R_{t}^{*})E_{t}(\epsilon_{t})-\gamma F_{t}^{2} 
\mu_t (1+R_{t}^{*})^2V_{t}(\epsilon_{t}).
\end{equation*}
\begin{equation*}
\Rightarrow \ \ \ \frac{\partial E_{t}(F_{t+1})}{\partial\mu_{t}^*}=
- F_{t}(1+R_{t}^{D})+F_{t}(1+R_{t}^{*})E_{t}(\epsilon_{t}).
\end{equation*}
\begin{equation*}
\Rightarrow \ \ \ E_t(F_{t+1}^2)=V_{t}(F_{t+1})+(E_t(F_{t+1}))^2 .
\end{equation*}
Now,
\begin{equation*}
\frac{\partial 
(E_{t}(F_{t+1}))^2}{\partial\mu_{t}^*}=2E_t(F_{t+1})\frac{\partial 
E_{t}(F_{t+1})}{\partial\mu_{t}^*}
\end{equation*}
and,
\begin{equation*}
V_{t}(F_{t+1})= \mu_t^2(1+R_t^*)^2F_t^2V_t(\epsilon_t) .
\end{equation*}
Thus,
\begin{equation*}
\frac{\partial V_{t}(F_{t+1})}{\partial\mu_{t}^*}=2 \mu_t 
(1+R_t^*)^2F_t^2V_{t}(\epsilon_t)
\end{equation*}
\begin{equation*}
\Rightarrow \ \ \
\frac{\partial M_{t}^{D}}{\partial\mu_{t}^*}|\mu_{t}^*=\frac{\partial 
\Omega_{t}^{D}}{\partial\mu_{t}^*}+B^{D}b\frac{\partial 
E_{t}(F_{t+1})}{\partial\mu_{t}^*}+B^{D}c\frac{\partial 
V_{t}(F_{t+1})}{\partial\mu_{t}^*}+B^{D}c\frac{\partial 
(E_{t}(F_{t+1}))^2}{\partial\mu_{t}^*} .
\end{equation*}
This implies,
\begin{eqnarray}
\label{focmu}
0 &=& -F_{t}(1+R_{t}^{D})+F_{t}(1+R_{t}^{*})E_{t}(\epsilon_{t})-\gamma 
F_{t}^{2} \mu_t (1+R_{t}^{*})^2V_{t}(\epsilon_{t}) \nn\\
&& + B^{D}b[-F_{t}(1+R_{t}^{D})+F_{t}(1+R_{t}^{*})E_{t}(\epsilon_{t})]
+ B^{D}c[2 \mu_t (1+R_t^*)^2F_t^2V_t(\epsilon_t)] \nn\\
&& + 2B^{D}c[F_{t}[(1-\mu_{t}) (1+R_{t}^{D})+ \mu_t 
(1+R_{t}^{*})E_t(\epsilon_{t})]-r_{tk}^D] \nn\\
&& \times [-F_{t}(1+R_{t}^{D})+F_{t}(1+R_{t}^{*})E_t(\epsilon_{t})] . \nn\\
\end{eqnarray}
Rearranging,
\begin{eqnarray}
\label{eq18a}
\mu_t^* &=& \frac{F_t Z_t[(1+B^{D}b)+2B^{D}cF_t(1+R_t^D)-2B^{D}cr_{tk}^D]}
{F_t^2[\gamma (1+R_t^*)^2V_t(\epsilon_t)-
2B^{D}c(Z_t^2+(1+R_t^*)^2V_t(\epsilon_t))]} \nn\\
&=& \frac{Z_t[(1+B^{D}b)+2B^{D}cF_t(1+R_t^D)-2B^{D}cr_{tk}^D]}
{F_t[\gamma (1+R_t^*)^2V_t(\epsilon_t)-
2B^{D}c(Z_t^2+(1+R_t^*)^2V_t(\epsilon_t))]}
\end{eqnarray}
with
\begin{equation}
Z_{t}=[(1+R_{t}^{*})E_{t}(\epsilon_{t})-(1+R_{t}^{D})]
\end{equation}
and
\begin{equation}
r_{tk}^D=(1+r_{t})K_t^D-K_{t+1}^D ,
\end{equation}
where $b$ and $c$ are determined as given below.
Let $\mu_{t}^{*}$be the optimum portfolio, then the optimum portfolio 
satisfies
$a+bF_{t}+cF_{t}^{2}=\Omega_{t}^{D}(F_{t},\mu_{t}^{*})+B^{D}E_t(V(F_{t+1})(\mu_{t}^{*}))$ 
from equation \eqref{eq28}
This equation holds for every $F_t$, so $a$, $b$, $c$ are determined by 
comparing coefficients 
of $F_t$ , $F_t^2$ and constant on both the sides of the equation. This 
implies
\begin{eqnarray}
\label{eqabc}
\hspace{-25pt}
 a+bF_{t}+cF_{t}^{2} &=& F_{t}[(1-
\mu_{t}^{*})(1+R_{t}^{D})+\mu_t^{*}(1+R_{t}^{*})E_t(\epsilon_{t})] - 
\frac{\gamma}{2} F_t^{2}{\mu_{t}^{*}}^{2}(1+R_t^*)^2V_t(\epsilon) + B^{D}a 
\nn\\
&& + B^{D}b[F_{t}[(1-\mu_{t}^{*})(1+R_{t}^{D}) 
 + \mu_t^{*}(1+R_{t}^{*})E_t(\epsilon_{t})]-(1+r_{t}^{D})K_{t}^{D}+K_{t+1}^D] 
\nn\\
&& + B^{D}c[(E_t(F_{t+1}))^2+V_{t}(F_{t+1})].
\end{eqnarray}
This is so because $E_t(F_{t+1}^2)=(E_t(F_{t+1}))^2+V_t(F_{t+1})$
Comparing the coefficients of $F_{t}^2$ on both the sides of the equation, 
the following 
equality is obtained:

\begin{eqnarray}
\label{eqc0}
c = (1+R_t^*)^2 V_t(\ep_t) \left(\frac{2 B^D c(1+R^D_t) Z_t}{\mu_{denom}}\right)^2 (-(\gamma/2) + B^D c) 
+ (1 + R^D)^2 B^D c + \frac{(2 B^D c(1+R^D_t) Z_t)^2}{\mu_{denom}} , \nn\\
\end{eqnarray}
where $\mu_{denom} = (\gamma - 2 B^D c) (1+R_t^*)^2 V_t(\ep_t) - 2 B^D c Z_t^2$. This yields,
\begin{eqnarray}
\label{eqc1}
\mu_{denom}^2 c (1 - B^D (1 + R_t^D)^2) &=& 
(2 B^D c(1+R^D_t) Z_t)^2 [\mu_{denom} - (1/2)(\gamma - 2 B^D c) (1+R_t^*)^2 V_t(\ep_t)] \nn\\
& \Rightarrow & \nn\\
(A - c E)^2 (1 - T) &=& 2 c Z_t^2 B^D T (A - c E_1) \nn\\
& \Rightarrow & (\mbox{ as } c \neq 0, \mbox{ otherwise the value function would be linear }\nn\\
(A^2 + c^2 E^2 - 2 c A E) (1 - T) &=& 2 c Z_t^2 B^D T (A - c E_1) \nn\\
& \Rightarrow & \nn\\
0 &=& c^2 (E^2 (1-T) +  2 Z_t^2 B^D T E_1) - 2 c (A E (1 - T) + Z_t^2 B^D T A) + A^2(1 - T)  \nn\\
\end{eqnarray}
where
$A = \gamma (1+R_t^*)^2 V_t(\ep_t)$, \ $E=2 B^D (Z_t^2 + (1+R_t^*)^2 V_t(\ep_t)) = 2 B^D (Z_t^2 
+ \frac{A}{\gamma})$, \ 
$T=B^D(1+R_t^D)^2$ \ and \ $E_1=2B^D (Z_t^2 + \frac{A}{2\gamma})$. Hence,
Note, since $A$, $E$ and $T$ all are positive, the coefficient of $c$ in \eqref{eqc1} is negative.
Therefore, if $0 < T \le 1$, then all solution of $c$ are positive real, as the discriminant
is positive,
since for $0 < T \le 1$, \ 
$A^2 E^2 (1 - T)^2 = A^2 E^2 (1 - T)^2$, \ $2 B^D Z_t^2 A^2 E T (1 - T) \ge 2 B^D Z_t^2 A^2 E_1 T (1 - T)$ \ and 
\ $(B^D)^2 Z_t^4 A^2 T^2 > 0$.
Hence, the necessary condition is $T > 1$. But then the discriminant may not always be positive
as the 2nd condition may not always hold and the coefficient of $c^2$ may also be negative.
So, the main idea would be to choose $B^D$, such that $T > 1$, i.e, coefficient of $c^0$ is negative,
but the coefficient of $c^2$ is positive.
For $T > 1$, when coefficient of $c^2$ is positive
then there exists exactly one solution of $c$ which is negative. 
Therefore $c < 0$ if and only if 
$B^D (1+R_t^D)^2 > 1$ \ and \
\begin{eqnarray}
\label{eqcns}(1+R_t^D)^2
(Z_t^2 + (1+R_t^*)^2 V_t(\ep_t))^2  &>& T [(Z_t^2 + (1+R_t^*)^2 V_t(\ep_t))^2 
- Z_t^2  (Z_t^2 + (1/2)(1+R_t^*)^2 V_t(\ep_t))^2] \nn\\
&=& T [(3/2) Z_t^2 (1+R_t^*)^2 V_t(\ep_t) + ((1+R_t^*)^2 V_t(\ep_t))^2  \nn\\
&\Leftrightarrow & B^D (1+R_t^D)^2  < \frac{(Z_t^2 + (1+R_t^*)^2 V_t(\ep_t))^2 }
{(3/2) Z_t^2 (1+R_t^*)^2 V_t(\ep_t) + ((1+R_t^*)^2 V_t(\ep_t))^2} . \nn\\
\end{eqnarray}

\bigskip
For the coefficient of $F_t$,
\begin{eqnarray}
\label{eqb}
b &=& 2 B^D c \frac{Z_t^2  (1+R_t^D)}{\mu_{denom}} + (1+R_t^D) - \frac{\gamma}{2} (1+R_t^*)^2 V_t(\ep_t) 
4 B^D c (1 + R_t^D) \frac{Z_t^2}{\mu_{denom}^2} [(1+B^D b) - 2 B^D c r_{tk}^D ] \nn\\
&& + B^D b[(1+R^D) + \frac{Z_t^2}{\mu_{denom}^2} 2 B^D c (1 + R_t^D)] 
+ B^D c[\frac{2 Z_t^2  (1+R_t^D)}{\mu_{denom}} [(1+B^D b) - 2 B^D c r_{tk}^D ] \nn\\
&& + \frac{2 Z_t^4 (1+R_t^D)}{\mu_{denom}^2} 2B^D c [(1+B^D b) - 2 B^D c r_{tk}^D ] 
- 2 r_{tk}^D [(1+R_t^D) + 2 B^D c \frac{Z_t^2  (1+R_t^D)}{\mu_{denom}} ]] \nn\\
&& + B^D c (1+R_t^*)^2 V_t(\ep_t) 
4 B^D c (1 + R_t^D) \frac{Z_t^2}{\mu_{denom}^2} [(1+B^D b) - 2 B^D c r_{tk}^D ] \nn\\
&=& (1 + R_t^D) (1 + 2 B^D c\frac{Z_t^2}{\mu_{denom}}) [(1+B^D b) - 2 B^D c r_{tk}^D ] \nn\\
&& + 2 B^D c (1 + R_t^D) \frac{Z_t^2}{\mu_{denom}} 
(1 + 2 B^D c\frac{Z_t^2}{\mu_{denom}}) [(1+B^D b) - 2 B^D c r_{tk}^D ] \nn\\
&& + (1+R_t^*)^2 V_t(\ep_t) 4 B^D c\frac{Z_t^2}{\mu_{denom}^2}) 
[(1+B^D b) - 2 B^D c r_{tk}^D ][B^D c - \frac{\gamma}{2}] \nn\\
&=& [(1+B^D b) - 2 B^D c r_{tk}^D ] (1 + R_t^D) (1 + 2 B^D c\frac{Z_t^2}{\mu_{denom}})
(1 + 2 B^D c\frac{Z_t^2}{\mu_{denom}}) \nn\\
&& + [(1+B^D b) - 2 B^D c r_{tk}^D ] 
(1+R_t^*)^2 V_t(\ep_t) 4 B^D c(1+R_t^D)\frac{Z_t^2}{\mu_{denom}^2}) [B^D c - \frac{\gamma}{2}] \nn\\
&=& [(1+B^D b) - 2 B^D c r_{tk}^D ] [(1 + R_t^D) (1 + 2 B^D c\frac{Z_t^2}{\mu_{denom}})^2 \nn\\
&& - (\frac{\gamma}{2} - B^D c) (1+R_t^*)^2 V_t(\ep_t) 4 B^D c(1+R_t^D)\frac{Z_t^2}{\mu_{denom}^2}] .
\end{eqnarray}
This implies,
\begin{eqnarray}
\label{eqb1}
b(1 - B^D L) &=& (1 - 2 B^D c r_{tk}^D) L \nn\\
& \Rightarrow & \nn\\
b &=& \frac{(1 - 2 B^D c r_{tk}^D) L}{1 - B^D L} ,
\end{eqnarray}
where
\begin{eqnarray}
\label{eqL1}
L &=& (1 + R_t^D) \left(1 + \frac{2 B^D c Z_t^2}{\mu_{denom}}\right)^2
- (\gamma - 2 B^D c) (1+R_t^*)^2 V_t(\ep_t) (1+R_t^D)\frac{2 B^D c Z_t^2}{\mu_{denom}^2} \nn\\
&=& (1 + R_t^D) \left(1 -  \frac{- 2 B^D c Z_t^2}{\mu_{denom}}\right)^2
+ (1+R_t^D) \frac{(\gamma - 2 B^D c) (1+R_t^*)^2 V_t(\ep_t)}{\mu_{denom}} 
\frac{- 2 B^D c Z_t^2}{\mu_{denom}} \nn\\
&=& (1 + R_t^D) \left(1 -  \frac{- 2 B^D c Z_t^2}{\mu_{denom}}\right)^2
+ (1+R_t^D) \left(1- \frac{- 2 B^D c Z_t^2}{\mu_{denom}}\right)
\frac{- 2 B^D c Z_t^2}{\mu_{denom}} , 
\end{eqnarray}
since $\mu_{denom} -  (\gamma - 2 B^D c) (1+R_t^*)^2 V_t(\ep_t) = - 2 B^D c Z_t^2$.
Hence,
\begin{eqnarray}
\label{eqL2}
L &=& (1 + R_t^D) \left(1 -  \frac{- 2 B^D c Z_t^2}{\mu_{denom}}\right) 
\left[1 -  \frac{- 2 B^D c Z_t^2}{\mu_{denom}} + \frac{- 2 B^D c Z_t^2}{\mu_{denom}} \right] \nn\\
&=& (1+R_t^D) \left(1- \frac{- 2 B^D c Z_t^2}{\mu_{denom}}\right) . 
\end{eqnarray}
This immediately implies that $B^D L < 1$, as $z$ is negative and $\frac{- 2 B^D c Z_t^2}{\mu_{denom}} < 1$.
Also, note
$$1 + B^D b = \frac{(1 -  2 B^D c r_{tk}^D (B^D L))}{1 - B^D L} .
$$

\bigskip
Positivity of the $\mu_t^*$ may be found from the condition as given below.
\begin{eqnarray}
\label{eqmupos}
\mu_t^* > 0 
& \Leftrightarrow & \frac{1-2B^D c r_{tk}^D (B^D L)}{1 - B^D L} + 2B^D c[F_t (1+R_t^D) - r_{tk}^D] > 0 \nn\\
& \Leftrightarrow & 1-2B^D c r_{tk}^D (B^D L) + 2B^D c (1 - B^D L) [F_t (1+R_t^D) - r_{tk}^D] > 0 \nn\\
& \Leftrightarrow & 1-2B^D c r_{tk}^D + 2B^D c (1 - B^D L) [F_t (1+R_t^D)] > 0 \nn\\
& \Leftrightarrow & 2B^D c [r_{tk}^D - (1 - B^D L) [F_t (1+R_t^D)]] < 1 . 
\end{eqnarray}
If $r_{tk}^D > (1 - B^D L) [F_t (1+R_t^D)]$, then it always holds as $c < 0$. This should be
the case, as it means $F_t$, is tool small compared to $r_{tk}^D$. On the other hand,
if $r_{tk}^D < (1 - B^D L) [F_t (1+R_t^D)]$, which is often the case, then there exists
$\gamma_1 > 0$ such that, $\gamma < \gamma_1$ would imply the last condition, as $c$ is linear
in $\gamma$. This means if the loan do not have too high a risk premium then $\mu_t^* > 0$.

Again, the condition for $\mu_t^* < 1$ may be found from below:
\begin{eqnarray}
\mu_t^* < 1 
& \Leftrightarrow & Z_t \left(\frac{1-2B^D c r_{tk}^D (B^D L)}{1 - B^D L} 
+ 2B^D c[F_t (1+R_t^D) - r_{tk}^D]\right) < \mu_{denom} F_t \nn\\
& \Leftrightarrow & Z_t \left((1-2B^D c r_{tk}^D (B^D L)) 
+ 2B^D c (1 - B^D L) [F_t (1+R_t^D) - r_{tk}^D]\right) < \mu_{denom} F_t (1 - B^D L)  \nn\\
& \Leftrightarrow & Z_t \left((1-2B^D c r_{tk}^D) 
+ 2B^D c (1 - B^D L) [F_t (1+R_t^D)]\right) < \mu_{denom} F_t (1 - B^D L)  \nn\\
& \Leftrightarrow & Z_t < F_t (1 - B^D L) \mu_{denom} 
+ 2B^D c Z_t [r_{tk}^D - (1 - B^D L) F_t (1+R_t^D)] .
\end{eqnarray}
Since $\mu_{denom}$ is linear in $\gamma$, as $c$ is, from \eqref{eqmupos} it is clear that
there exists $0 < \gamma_0 < \gamma_1$, such that for $\gamma > \gamma_0$ the last condtion
holds. This means if the riskpremium is to low then there is always a possibility of over-subscription
of lending.

Hence the proof of existence of $\mu^* \in (0, 1)$ \ for \ $\gamma \in (\gamma_0, \gamma_1)$.
Upper bound $\gamma_1$ may be infinity if $F_t$ remains too small. 

\section{Appendix B}
\label{app2}
\setcounter{equation}{0}

{\bf Proof of Theorem \ref{th2}:} \                                                     
To solve the maximization problem, denote 
\begin{equation*}
M_{t}^{U}=\Omega_{t}^{U}+B^{U}E_t(V^{U}(G_{t+1})).
\end{equation*} 
Then
\begin{equation*}
V^{U}(G_{t})=max_{\lambda_{t}}M_{t}^{U}.
\end{equation*}
First order condition for maximization implies,
\begin{equation}
\frac{\partial 
M_{t}^{U}}{\partial\lambda_{t}}{\mid}_{\lambda_t=\lambda_{t}^{*}}=0
\end{equation}
Now,
\begin{equation*}
\frac{\partial \Omega_{t}^{U}}{\partial\lambda_{t}}
=K_{t}^{U}A_t+\beta \lambda_t 
K_{t}^{U^{2}}(1+R_{t}^{*})^{2}V_{t}(\frac{e_{t+1}}{e_{t}})
\end{equation*}
where,
\begin{equation*}
A_t=(1+r_t^U)-(1+R_t^*)E_t(\frac{e_{t+1}}{e_t}) .
\end{equation*}
Further,
\begin{equation*}
E_{t}(G_{t+1})=E_t(\eta_{t})(1+R_{t}^{U})G_{t}-K_{t}^{U}[(1-
\lambda_{t})(1+r_{t}^{U})+\lambda_{t}(1+R_{t}^{*})E_{t}(\frac{e_{t+1}}{e_{t}}
)]+K_{t+1}^{U}.
\end{equation*}
So,
\begin{equation*}
\frac{\partial E_{t}(G_{t+1})}{\partial\lambda_{t}}=K_t^U A_t ,
\end{equation*}
and,
$E_t(G_{t+1}^2)=(E_t(G_{t+1}))^2+V_{t}(G_{t+1})$. 

Now,
\begin{equation*}
\frac{\partial 
(E_{t}(G_{t+1})^2}{\partial\lambda_{t}}=2E_t(G_{t+1})\frac{\partial 
E_{t}(G_{t+1}}{\partial\lambda_{t}}
\end{equation*}
\begin{equation*}
\Rightarrow \ \ \
V_t(G_{t+1})=K_t^{U^2}\lambda_t^2(1+R_t^*)^2V_{t}(\frac{e_{t+1}}{e_t})+V_t(\e
ta_t)G_t^2(1+R_t^U)^2.
\end{equation*}
Thus
\begin{equation*}
\frac{\partial 
V_{t}(G_{t+1})}{\partial\lambda_{t}}=2K_t^{U^2}\lambda_t(1+R_t^*)^2 
V_{t}(\frac{e_{t+1}}{e_t}).
\end{equation*}
\begin{equation*}
\Rightarrow \ \ \
\frac{\partial M_{t}^{U}}{\partial\lambda_{t}}=\frac{\partial 
\Omega_{t}^{U}}{\partial\lambda_{t}}+B^{U}y\frac{\partial 
E_{t}(G_{t+1})}{\partial\lambda_{t}}+B^{U}z\frac{\partial 
V_{t}(G_{t+1})}{\partial\lambda_{t}}+B^{U}z\frac{\partial 
(E_{t}(G_{t+1}))^2}{\partial\lambda_{t}}.
\end{equation*}
This implies,
\begin{eqnarray*}
\hspace{-25pt}
0 &=& K_{t}^{U}A_{t} - \beta \lambda_t 
K_{t}^{U^{2}}(1+R_{t}^{*})^{2}V_{t}(\frac{e_{t+1}}{e_{t}}) 
+ B^{U}y(A_tK_t^U)+ 
B^{U}z[2K_t^{U^2}\lambda_t(1+R_t^*)^2V_{t}(\frac{e_{t+1}}{e_t})] \nn\\
&& + 2B^{U}z(A_tK_t^U)[(E_t(\eta_{t})(1+R_{t}^{U})G_{t}-K_{t}^{U}[(1-
\lambda_{t})(1+r_{t}^{U})+\lambda_{t}(1+R_{t}^{*})E_{t}(\frac{e_{t+1}}{e_{t}}
)]+K_{t+1}^U] . \nn\\
\end{eqnarray*}
Rearranging,
\begin{eqnarray}
\label{eq40}
\lambda_t^* &=& \frac{K_t^{U}A_t[(1+B^{U}y)
+2B^{U}z(1+R_t^U)E_t(\eta_t)G_t-2B^{U}z r_{tk}^U]}
{K_t^{U^2}[(1+R_t^*)^2V_t(\frac{e_{t+1}}{e_t})
(\beta-2B^{U}z)-2B^{U}zA_t^2]} \nn\\
&=& \frac{A_t[(1+B^{U}y)
+2B^{U}z(1+R_t^U)E_t(\eta_t)G_t-2B^{U}z r_{tk}^U]}
{K_t^{U}[(1+R_t^*)^2V_t(\frac{e_{t+1}}{e_t})
(\beta-2B^{U}z)-2B^{U}zA_t^2]}
\end{eqnarray}
where
\begin{equation*}
r_{tk}^U=K_t^U(1+r_t^U)- K_{t+1}^U
\end{equation*}
$x,y,z$ are determined as in below.
If $\lambda_{t}^{*}$ is optimal then 
\begin{equation}
x+yG_{t}+zG_{t}^{2}=\Omega_{t}^{U}(G_{t},\lambda_{t}^{*})+B^{U}E_t(V^{U}(G_{t
+1})).
\end{equation}
$x, y, z$ are obtained by comparing the coefficients of $G_t, G_t^2$ and 
constants on both the sides of the equation. Substituting for value function 
$E_t(V^{U}(G_{t+1}))$ we get
$z=-\frac{\beta}{2}(1+R_t^U)^2V_t(\eta_t)+B^{U}z$(coeff of $G_{t}^2$ in 
$(E_t(G_{t+1}))^2+B^{U}z$(coeff of $G_t^2$ in $V_{t}(G_{t+1}))$,
i.e.,

for the coefficient of $G_t^2$,
\begin{eqnarray}
z &=& - \frac{\beta}{2} [(1 + R_t^U)^2 V_t(\eta_t) 
+ (1 + R_t^*)^2 V_t(\frac{e_{t+1}}{e_t}) \frac{A_t^2}{\lambda_{denom}^2} 
4 (B^U)^2 z^2 (1 + R_t^U)^2 (E_t(\eta_t))^2] \nn\\
&& + B^U z[ (1 + R_t^U)^2 (E_t(\eta_t))^2 + \frac{A_t^4}{\lambda_{denom}^2} 
4 (B^U)^2 z^2 (1 + R_t^U)^2 (E_t(\eta_t))^2 
+ \frac{A_t^2}{\lambda_{denom}} 4 B^U z (1 + R_t^U)^2 (E_t(\eta_t))^2 ] \nn\\
&& + B^U z [ (1 + R_t^U)^2 V_t(\eta_t) 
+ (1 + R_t^*)^2 V_t(\frac{e_{t+1}}{e_t}) \frac{A_t^2}{\lambda_{denom}^2} 
4 (B^U)^2 z^2 (1 + R_t^U)^2 (E_t(\eta_t))^2 ] , 
\end{eqnarray}
where $\lambda_{denom} = [(1+R_t^*)^2V_t(\frac{e_{t+1}}{e_t}) (\beta-2B^{U}z)-2B^{U}zA_t^2] . $
This implies,
\begin{eqnarray}
\label{solz}
z \lambda_{denom}^2 &=& \lambda_{denom}^2 (1 + R_t^U)^2 V_t(\eta_t) [B^U z - \frac{\beta}{2}] \nn\\
&& + (1 + R_t^*)^2 V_t(\frac{e_{t+1}}{e_t}) A_t^2 4 (B^U)^2 z^2 (1 + R_t^U)^2 (E_t(\eta_t))^2 
[B^U z - \frac{\beta}{2}] \nn\\
&& + B^U z (1 + R_t^U)^2 (E_t(\eta_t))^2 [ \lambda_{denom}^2 + A_t^4 4 (B^U)^2 z^2 + \lambda_{denom} A_t^2 4 B^U z ] . 
\end{eqnarray}

Note that, if we bring everything to the left and equate it to zero, we obtain, coefficient of $z^3$ as,
\begin{eqnarray}
\label{solz1}
&& (2B^U)^2 H^2 - (2B^U)^2 H^2 B^U (1 + R_t^U)^2 V_t(\eta_t) - (2B^U)^2 (1 + R_t^*)^2 
V_t(\frac{e_{t+1}}{e_t}) A_t^2 (1 + R_t^U)^2 (E_t(\eta_t))^2 B^U \nn\\
&& - (2B^U)^2 (1 + R_t^U)^2 (E_t(\eta_t))^2 H^2 B^U - (2B^U)^2 A_t^4 (1 + R_t^U)^2 (E_t(\eta_t))^2 B^U \nn\\
&& - (2B^U)^2 A_t^2 (1 + R_t^U)^2 (E_t(\eta_t))^2 (-2B^U) H \nn\\
&=& (2B^U)^2 H^2 [1 - B^U (1 + R_t^U)^2 V_t(\eta_t) - B^U (1 + R_t^U)^2 (E_t(\eta_t))^2 ] 
+ (2B^U)^2 A_t^2 (1 + R_t^U)^2 (E_t(\eta_t))^2 B^U H , \nn\\
\end{eqnarray}
where $H = (A_t^2 + (1 + R_t^*)^2 V_t(\frac{e_{t+1}}{e_t}))$.
This coefficient is positive if $B^U (1 + R_t^U)^2 [ V_t(\eta_t) + (E_t(\eta_t))^2] < 1$, i.e.,
$B^U (1 + R_t^U)^2 E_t(\eta_t^2) < 1$.

Now, the coefficient of $z^0$, in the similar expression as above is,
$$\frac{\beta}{2} (1 + R_t^U)^2 V_t(\eta_t) [(1 + R_t^*)^2 V_t(\frac{e_{t+1}}{e_t})\beta]^2 ,$$
which is positive.
Since the coefficient of $z^3$ and $z^0$ are both positive it guarantees a negative solution of $z$.

If we can get a condition which gives the coefficient of $z$ to be negative then the Descartes' rule
of sign says there is at most one negative root.

Now the coefficient of $z$ is,
\begin{eqnarray}
&& [(1 + R_t^*)^2 V_t(\frac{e_{t+1}}{e_t})\beta]^2 - B^U (1 + R_t^U)^2 V_t(\eta_t) 
[(1 + R_t^*)^2 V_t(\frac{e_{t+1}}{e_t})\beta]^2 \nn\\
&& + (\beta/2) (1 + R_t^U)^2 V_t(\eta_t) (-4 B^U) H [(1 + R_t^*)^2 V_t(\frac{e_{t+1}}{e_t})\beta] \nn\\
&& - B^U (1 + R_t^U)^2 (E_t(\eta_t))^2 [(1 + R_t^*)^2 V_t(\frac{e_{t+1}}{e_t})\beta]^2 \nn\\
&=& [(1 + R_t^*)^2 V_t(\frac{e_{t+1}}{e_t})\beta]^2 [1 - 3 B^U (1 + R_t^U)^2 V_t(\eta_t)
- B^U (1 + R_t^U)^2 (E_t(\eta_t))^2 ] \nn\\
&& - 2 \beta^2 B^U (1 + R_t^U)^2 V_t(\eta_t) A_t^2 (1 + R_t^*)^2 V_t(\frac{e_{t+1}}{e_t}) .
\end{eqnarray}
This expression is negative if,
\begin{equation}
\label{eqcoefz}
3B^U (1 + R_t^U)^2 V_t(\eta_t) + B^U (1 + R_t^U)^2 (E_t(\eta_t))^2 > 1 . 
\end{equation}

Thus there exists exactly one negative solution $z$ if and only if
\begin{eqnarray}
\label{eqnbu}
 \frac{1}{3 (1 + R_t^U)^2 V_t(\eta_t) +  (1 + R_t^U)^2 (E_t(\eta_t))^2} < B^U
< \frac{1}{(1 + R_t^U)^2 [ V_t(\eta_t) + (E_t(\eta_t))^2]} .
\end{eqnarray}


For the coefficient of $z^2$, \ in \ $a_0 + a_1 z + a_2 z^2 + a_3 z^3 = 0$ is
\begin{eqnarray}
&& - 4B^U H (\beta (1+R_t^*)^2  V_t(\frac{e_{t+1}}{e_t}) 
- B^U (1+R_t^U)^2 [(E_t(\eta_t))^2 + V_t(\eta_t)] [(-4B^U) H (\beta (1+R_t^*)^2  V_t(\frac{e_{t+1}}{e_t}) ] \nn\\
&& + (\beta/2) (1 + R_t^U)^2 V_t(\eta_t) (-2B^U)^2 H^2 
+ (\beta/2) (1 + R_t^U)^2 (E_t(\eta_t))^2  (1+R_t^*)^2  V_t(\frac{e_{t+1}}{e_t}) A_t^2 4 (B^U)^2 \nn\\
&& - B^U (1 + R_t^U)^2 (E_t(\eta_t))^2 A_t^2 4 B^U (1+R_t^*)^2  V_t(\frac{e_{t+1}}{e_t}) \beta \nn\\
&=& - 4B^U H (\beta (1+R_t^*)^2  V_t(\frac{e_{t+1}}{e_t}) 
+ 4 (B^U)^2 (1+R_t^U)^2 E_t(\eta_t^2)  H (\beta (1+R_t^*)^2  V_t(\frac{e_{t+1}}{e_t}) \nn\\
&& + (\beta/2) 4(B^U)^2 (1+R_t^U)^2 V_t(\eta_t) H^2 
- (\beta/2) 4 (B^U)^2 (1+R_t^U)^2 (E_t(\eta_t))^2 (1+R_t^*)^2  V_t(\frac{e_{t+1}}{e_t}) A_t^2 .
\end{eqnarray}

\bigskip
For the coefficient of $G_t$,
\begin{eqnarray}
y &=& (1+R_t^U) E_t (\eta_t) + \frac{A_t^2}{\lambda_{denom}} 2 B^U z (1+R_t^U) E_t (\eta_t)   \nn\\
&& - \frac{\beta}{2} (1 + R_t^*)^2 V_t(\frac{e_{t+1}}{e_t}) [ \frac{2 A_t^2}{\lambda_{denom}^2} 2 B^U z (1+R_t^U) E_t (\eta_t)
[(1 + B^U y) -2 B^U z r_{tk}^U ]] \nn\\
&& + B^U y [(1+R_t^U) E_t (\eta_t) + \frac{A_t^2}{\lambda_{denom}} 2 B^U z (1+R_t^U) E_t (\eta_t) ] \nn\\
&& + B^U z [2 (1+R_t^U) E_t (\eta_t) 
[\frac{A_t^2}{\lambda_{denom}} [(1 + B^U y) -2 B^U z r_{tk}^U ] - r_{tk}^U] \nn\\
&& + \frac{A_t^2}{\lambda_{denom}} 4 B^U z (1+R_t^U) E_t (\eta_t) 
[\frac{A_t^2}{\lambda_{denom}} [(1 + B^U y) -2 B^U z r_{tk}^U ] - r_{tk}^U] ] \nn\\
&& + B^U z (1 + R_t^*)^2 V_t(\frac{e_{t+1}}{e_t}) 
[\frac{A_t^2}{\lambda_{denom}^2} 4 B^U z (1+R_t^U) E_t (\eta_t) [(1 + B^U y) -2 B^U z r_{tk}^U ] ] . \nn\\ 
\end{eqnarray}
This implies
\begin{eqnarray}
y &=& (1+R_t^U) E_t (\eta_t) [ 1 + \frac{2 B^U z A_t^2}{\lambda_{denom}} ]   \nn\\
&& + B^U y [(1+R_t^U) E_t (\eta_t)] [1 + \frac{2 B^U z A_t^2}{\lambda_{denom}} ] \nn\\
&& + B^U z [2 (1+R_t^U) E_t (\eta_t) 
[\frac{A_t^2}{\lambda_{denom}} [(1 + B^U y) -2 B^U z r_{tk}^U ] - r_{tk}^U] ] 
[1 + \frac{2 B^U z A_t^2}{\lambda_{denom}} ] \nn\\
&& + (1 + R_t^*)^2 V_t(\frac{e_{t+1}}{e_t}) 
[\frac{2 B^U z A_t^2}{\lambda_{denom}^2}  (1+R_t^U) E_t (\eta_t) [(1 + B^U y) -2 B^U z r_{tk}^U ] ] 
[2 B^U z - \beta] . \nn\\ 
\end{eqnarray}
Thus,
\begin{eqnarray}
y &=& (1+R_t^U) E_t (\eta_t) [ 1 + \frac{2 B^U z A_t^2}{\lambda_{denom}} ]   
+ B^U y [(1+R_t^U) E_t (\eta_t)] [1 + \frac{2 B^U z A_t^2}{\lambda_{denom}} ] \nn\\
&& + B^U y [(1+R_t^U) E_t (\eta_t) 
[\frac{2 B^U z A_t^2}{\lambda_{denom}} ]
[1 + \frac{2 B^U z A_t^2}{\lambda_{denom}} ]] \nn\\
&& + B^U z [2 (1+R_t^U) E_t (\eta_t) 
[\frac{A_t^2}{\lambda_{denom}} [1 - 2 B^U z r_{tk}^U ] - r_{tk}^U] ] 
[1 + \frac{2 B^U z A_t^2}{\lambda_{denom}} ] \nn\\
&& + B^U y (1 + R_t^*)^2 V_t(\frac{e_{t+1}}{e_t}) 
[\frac{2 B^U z A_t^2}{\lambda_{denom}^2}  (1+R_t^U) E_t (\eta_t)  ] 
[2 B^U z - \beta] . \nn\\ 
&& + (1 + R_t^*)^2 V_t(\frac{e_{t+1}}{e_t}) 
[\frac{2 B^U z A_t^2}{\lambda_{denom}^2}  (1+R_t^U) E_t (\eta_t) [1 - 2 B^U z r_{tk}^U ] ] 
[2 B^U z - \beta] . \nn\\ 
\end{eqnarray}
Therefore,
\begin{eqnarray}
y &=& (1+R_t^U) E_t (\eta_t) [ 1 + \frac{2 B^U z A_t^2}{\lambda_{denom}} ]   
+ B^U y [(1+R_t^U) E_t (\eta_t)] [1 + \frac{2 B^U z A_t^2}{\lambda_{denom}} ]^2 \nn\\
&& + B^U z [2 (1+R_t^U) E_t (\eta_t) 
[\frac{A_t^2}{\lambda_{denom}} [1 - 2 B^U z r_{tk}^U ] - r_{tk}^U] ] 
[1 + \frac{2 B^U z A_t^2}{\lambda_{denom}} ] \nn\\
&& + B^U y (1 + R_t^*)^2 V_t(\frac{e_{t+1}}{e_t}) 
[\frac{2 B^U z A_t^2}{\lambda_{denom}^2}  (1+R_t^U) E_t (\eta_t)  ] 
[2 B^U z - \beta] . \nn\\ 
&& + (1 + R_t^*)^2 V_t(\frac{e_{t+1}}{e_t}) 
[\frac{2 B^U z A_t^2}{\lambda_{denom}^2}  (1+R_t^U) E_t (\eta_t) [1 - 2 B^U z r_{tk}^U ] ] 
[2 B^U z - \beta] . \nn\\ 
\end{eqnarray}
Hence,
\begin{eqnarray}
y [1 - J_1 + J_2]
&=& (1+R_t^U) E_t (\eta_t) [ 1 + \frac{2 B^U z A_t^2}{\lambda_{denom}} ]   
 \nn\\
&& + B^U z [2 (1+R_t^U) E_t (\eta_t) 
[\frac{A_t^2}{\lambda_{denom}} [1 - 2 B^U z r_{tk}^U ] - r_{tk}^U] ] 
[1 + \frac{2 B^U z A_t^2}{\lambda_{denom}} ] \nn\\ 
&& + (1 + R_t^*)^2 V_t(\frac{e_{t+1}}{e_t}) 
[\frac{2 B^U z A_t^2}{\lambda_{denom}^2}  (1+R_t^U) E_t (\eta_t) [1 - 2 B^U z r_{tk}^U ] ] 
[2 B^U z - \beta]  \nn\\ 
&=& (1+R_t^U) E_t (\eta_t) [ 1 - \frac{-2 B^U z A_t^2}{\lambda_{denom}} ]   
 \nn\\
&& + [(1+R_t^U) E_t (\eta_t)] 
\frac{2 B^U z A_t^2}{\lambda_{denom}} [1 - 2 B^U z r_{tk}^U ]
[1 - \frac{- 2 B^U z A_t^2}{\lambda_{denom}} ] \nn\\ 
&& + (1+R_t^U) E_t (\eta_t) (- 2 B^U z r_{tk}^U) [1 - \frac{- 2 B^U z A_t^2}{\lambda_{denom}} ] \nn\\ 
&& - (1+R_t^U) E_t (\eta_t) [1 - 2 B^U z r_{tk}^U ] 
[\frac{(\beta - 2 B^U z)(1 + R_t^*)^2 V_t(\frac{e_{t+1}}{e_t})}{\lambda_{denom}} ]
[\frac{2 B^U z A_t^2}{\lambda_{denom}}  ]  \nn\\ 
&=& (1+R_t^U) E_t (\eta_t) [ 1 - \frac{-2 B^U z A_t^2}{\lambda_{denom}} ] [1 - 2 B^U z r_{tk}^U]   \nn\\
&& + [(1+R_t^U) E_t (\eta_t)] 
\frac{2 B^U z A_t^2}{\lambda_{denom}} [1 - 2 B^U z r_{tk}^U ]
[1 - \frac{- 2 B^U z A_t^2}{\lambda_{denom}} ] \nn\\ 
&& - (1+R_t^U) E_t (\eta_t) [1 - 2 B^U z r_{tk}^U ] 
[1 - \frac{-2 B^U z A_t^2}{\lambda_{denom}}  ]  
[\frac{2 B^U z A_t^2}{\lambda_{denom}}  ]  \nn\\ 
&=& (1+R_t^U) E_t (\eta_t) [ 1 - \frac{-2 B^U z A_t^2}{\lambda_{denom}} ] [1 - 2 B^U z r_{tk}^U] , \nn\\
\end{eqnarray}
since
$\lambda_{denom} - (\beta - 2 B^U z)(1 + R_t^*)^2 V_t(\frac{e_{t+1}}{e_t}) = -2 B^U z A_t^2$,
where

$J_1 = B^U  [(1+R_t^U) E_t (\eta_t)] [1 + \frac{2 B^U z A_t^2}{\lambda_{denom}} ]^2$,
and

$J_2 = B^U (1 + R_t^*)^2 V_t(\frac{e_{t+1}}{e_t}) 
[\frac{2 B^U z A_t^2}{\lambda_{denom}^2}  (1+R_t^U) E_t (\eta_t)  ] 
[\beta - 2 B^U z] $.

As in $L$ \eqref{eqL1} and \eqref{eqL2}  in the computation of $y$,
the expression for $J_1 - J_2$ can be simplified further.
Note,
\begin{eqnarray}
J_1 - J_2 &=& B^U  [(1+R_t^U) E_t (\eta_t)] \left[\left(1 - \frac{-2 B^U z A_t^2}{\lambda_{denom}} \right)^2
+ \frac{(\beta - 2 B^U z) (1 + R_t^*)^2 V_t(\frac{e_{t+1}}{e_t})}{\lambda_{denom}}
\left(\frac{-2 B^U z A_t^2}{\lambda_{denom}}\right) \right] \nn\\
&=& B^U  [(1+R_t^U) E_t (\eta_t)] \left[\left(1 - \frac{-2 B^U z A_t^2}{\lambda_{denom}} \right)^2
+ \left(1 - \frac{-2 B^U z A_t^2}{\lambda_{denom}} \right)
\left(\frac{-2 B^U z A_t^2}{\lambda_{denom}}\right) \right] \nn\\
&=& B^U  [(1+R_t^U) E_t (\eta_t)] \left[\left(1 - \frac{-2 B^U z A_t^2}{\lambda_{denom}} \right)
\left[\left(1 - \frac{-2 B^U z A_t^2}{\lambda_{denom}} \right) + 
\left(\frac{-2 B^U z A_t^2}{\lambda_{denom}}\right) \right] \right] \nn\\
&=& B^U  [(1+R_t^U) E_t (\eta_t)] \left(1 - \frac{-2 B^U z A_t^2}{\lambda_{denom}} \right)  \nn\\
&=& B^U J ,
\end{eqnarray}
where $J = [(1+R_t^U) E_t (\eta_t)] \left(1 - \frac{-2 B^U z A_t^2}{\lambda_{denom}} \right)$.
Thus,
\begin{eqnarray}
\label{soly}
y &=& \frac{(1+R_t^U) E_t (\eta_t) [ 1 - \frac{-2 B^U z A_t^2}{\lambda_{denom}} ] [1 - 2 B^U z r_{tk}^U]}
{1 - B^U J} \nn\\
&=& \frac{J (1 - 2 B^U z r_{tk}^U)}{(1 - B^U J)} .
\end{eqnarray}

From this, $y$ can be found immediately once the $z$ is known.

Now \eqref{soly} further implies
$$1 + B^U y = \frac{(1 - 2 B^U z r_{tk}^U (B^U J))}{1 - B^U J} .
$$

\bigskip
Positivity of the $\lambda_t^*$ may be found from the condition as given below:
\begin{eqnarray}
\label{eqlampos}
\lambda_t^* > 0 
& \Leftrightarrow & \frac{1-2B^U z r_{tk}^U (B^U J)}{1-B^D J} + 2B^U z[G_t (1+R_t^U)E(\eta_t) - r_{tk}^U] > 0 \nn\\
& \Leftrightarrow & 1-2B^U z r_{tk}^U (B^U J) + 2B^U z (1-B^D J)[G_t (1+R_t^U)E(\eta_t) - r_{tk}^U] > 0 \nn\\
& \Leftrightarrow & 1-2B^U z r_{tk}^U + 2B^U z (1-B^D J)[G_t (1+R_t^U)E(\eta_t)] > 0 \nn\\
& \Leftrightarrow & 2B^U z [r_{tk}^U - (1-B^D J)[G_t (1+R_t^U)E(\eta_t)]] < 1 . 
\end{eqnarray}
If $r_{tk}^U > (1 - B^U J) [G_t (1+R_t^U)E(\eta_t)]$, then it always holds as $z < 0$. This should be
the case, as it means $G_t$, is tool small compared to $r_{tk}^D$ (or perhaps negative). 
On the other hand,
if $r_{tk}^U < (1 - B^U J) [G_t (1+R_t^U)E(\eta_t)]$, 
then there exists
$\beta_1 > 0$ such that, $\beta < \beta_1$ would imply the last condition, as $z$ is linear
in $\beta$. This means if the loan do not have too high a risk premium then $\lambda_t^* > 0$.

Since $\lambda_{denom}$ is linear in $\beta$, as $z$ is, from \eqref{eqlampos} it is clear that
there exists $0 < \beta_0 < \beta_1$, such that for $\beta > \beta_0$ the last condtion
holds. This means if the risk premium is too low then there is always a possibility of over-subscription
of borrowing.

Hence the proof of existence of $\lambda_t^* \in (0, 1)$ \ for \ $\beta \in (\beta_0, \beta_1)$.
Upper bound $\beta_1$ may be infinity if $G_t$ remains too small or negative. 

Similarly, the condition for $\lambda_t^* < 1$ may be found from below:
\begin{eqnarray}
\lambda_t^* < 1 
& \Leftrightarrow & A_t \left(\frac{1-2B^U z r_{tk}^U (B^U J)}{1 - B^U J} 
+ 2B^U z[G_t (1+R_t^U) E(\eta_t) - r_{tk}^U]\right) < \lambda_{denom} K_t^U \nn\\
& \Leftrightarrow & A_t \left( (1-2B^U z r_{tk}^U (B^U J))  
+ 2B^U z (1 - B^U J) [G_t (1+R_t^U) E(\eta_t) - r_{tk}^U]\right) < \lambda_{denom} K_t^U (1 - B^U J) \nn\\
& \Leftrightarrow & A_t \left( (1-2B^U z r_{tk}^U)  
+ 2B^U z (1 - B^U J) [G_t (1+R_t^U) E(\eta_t)]\right) < \lambda_{denom} K_t^U (1 - B^U J) \nn\\
& \Leftrightarrow & A_t <  (1 - B^U J) \lambda_{denom} K_t^U 
+ 2B^U z A_t \left(r_{tk}^U - (1 - B^U J) G_t (1+R_t^U) E(\eta_t)\right) . 
\end{eqnarray}
Since $\lambda_{denom}$ is linear in $\beta$, as $z$ is, from \eqref{eqlampos} it is clear that
there exists $0 < \beta_0 < \beta_1$, such that for $\beta > \beta_0$ the last condtion
holds. This means if the risk premium is too low then there is always a possibility of over-subscription
of borrowing.

Hence the proof of existence of $\lambda_t^* \in (0, 1)$ \ for \ $\beta \in (\beta_0, \beta_1)$.
Upper bound $\beta_1$ may be infinity if $G_t$ remains too small or negative.


Since it has multiple roots, we consider only those roots which belong to 
[0,1].\footnote
{uniroot command in rootSolve package in R is used for the root approximation 
in a specified 
interval.} \footnote{Here too, for some parametric configurations, existence 
of root in [0,1] 
is not guaranteed. $1 - \lambda_t^* <0$ $\Rightarrow \lambda_t^*>1$, 
excessive borrowing, generally 
happens whenever accumulated funds at time period t of a bank in the 
developing country 
$G_t$ is less than the total foreign loan taken by the bank.} 

\bigskip
{\bf Remark:}
Since $Z_t > 0$ and $A_t > 0$, are the economic requirement for the flow of Capital, we still have
the same restrictions on $R_t^*$, as 
$$\frac{(1+R_t^D)}{E_t(\epsilon_t)} < (1 + R_t^*) < \frac{(1+r_t^U)}{E_t(\frac{e_{t+1}}{e_t})} .$$

Besides,
hypothesis of Theorem 3 is also required, i.e.,
$$N_0^2 > 4 N_1 m^U K_t^U ,$$
to have a solution for $(R_t^*, e_t)$.

\section{Appendix C}
\label{app30}
\setcounter{equation}{0}

{\bf Proof of Theorem \ref{th3}:} Define
\begin{eqnarray}
\label{eqth31}
L_1(R_t^*,e_t) &=& m^D \mu_t^* F_t - \frac{m^U \lambda_t^* K_t^U}{e_t} \nn\\
\mbox{and} && \nn\\
L_2(R_t^*,e_t) &=& 
N_{1}e_{t}^2 - (N_{0} + C_{t-1})e_{t} + m^{U}\lambda_{t}^{*}K_{t}^{U} 
\end{eqnarray}
where
$C_{t-1}=\frac{m^{U}(1+R_{t-1}^{*})\lambda_{t-1}^{*}K_{t-1}^{U}}{e_{t-
1}^{*}}$.
Note, $L_2$ is a quadratic function of $e_{t}$ for every
$R_t^*$. Thus, equating $L_2$ with zero to get
\begin{eqnarray}
\label{eqet}
e_t = \frac{(N_{0} + C_{t-1}) \pm \sqrt{(N_{0} + C_{t-1})^2 - 4 
N_{1}m^{U}\lambda_{t}^{*}K_{t}^{U}}}{2 N_{1}} .
\end{eqnarray}
Since $N_{1}>0$ and $N_{0} + C_{t-1} > 0$, \ \eqref{eqet} has a unique 
positive solution if and only if
$(N_{0} + C_{t-1})^2 > 4 N_{1}m^{U}\lambda_{t}^{*}K_{t}^{U} $.
But the later holds if 
$N_{0}^2 > 4 N_{1}m^{U} K_{t}^{U}$
\ as $\lambda_t^* \in (0, 1)$ by Theorem \ref{th1}.

Again, from the assumption of Theorems \ref{th1}
and \ref{th2}, 
\begin{eqnarray}
\label{condR1}
\frac{(1+R_t^D)}{E_t(\epsilon_t)}<(1+R_t^*)<\frac{(1+r_t^U)}{E_t(\frac{e_{t+1
}}{e_t})} .
\end{eqnarray}
Thus, for every $e_t$, \
$L_1$ is positive if 
$(1+R_t^*)=\frac{(1+r_t^U)}{E_t(\frac{e_{t+1}}{e_t})}$ 
(i.e., $R_t^*=\frac{(1+r_t^U)}{E_t(\frac{e_{t+1}}{e_t})} - 1$) \ 
as $A_t=0$ implies $\lambda_t^*=0$.
Similarly, for every $e_t$, \ 
$L_1$ is negative if 
$(1+R_t^*)=\frac{(1+R_t^D)}{E_t(\epsilon_t)}$,
(i.e., $R_t^*=\frac{(1+R_t^D)}{E_t(\epsilon_t)} - 1 > 0$) \ 
as $Z_t=0$ implies $\mu_t^*=0$.
Also, for every $e_t$, \ $L_1$ is clearly an increasing
function of $R_t^*$. Hence there exist an unique
positive solution of $R_t^*$ for the equation
$L_1 = 0$ for every $e_t$. However, unique
positive solution $e_t$ is guaranteed by the above
assumption of the Theorem \ref{th3}. Hence the proof.

\section{Appendix D}
\label{app50}
\textbf{Algorithm of simulation}
\begin{itemize}
\item Start with initial values of endogenous variables $R_0^*=0.14$ and 
$e_0^*=40$ and other 
parametric values.
\item Calculate $\mu_0^*$ and $\lambda_0^*$ from $R_0^*$ using equations 
\eqref{eq18a} and 
\eqref{eq40} respectively.
\item To calculate $e_t$, $R_t^*$, $\lambda_t^*$ and $\mu_t^*$ at each t, 
fixed point iteration 
method is used. This is a version of majorize-minimization or an E-M 
algorithm in a dynamic context.
\item Step 1: Start with an initial value $\lambda_{tj}$, calculate $e_{tj}$ 
given $\lambda_{tj}^*$ 
using \eqref{eq36}
\item Step 2: $\lambda_{tj}^*=$ constant$\mu_{tj}^*\Rightarrow 
\lambda_{tj}^*=f(R_{tj}^*)$ from 
\eqref{eq18a}
\item Step 3: As $\lambda_{tj}^*$ is given(initial value), $R_{tj}^*$ is 
determined as the root of 
above polynomial.
\item Step 4: Obtaining $R_t^*$, update $\lambda_{t(j+1)}^*$ using 
\eqref{eq40},go back to step 1 and 
repeat the process until each variable converges.
\item Step 5: Having obtained all the exogenous variables at time period $t$, 
repeat the process to 
get those values at time period $t+1$.
 
\end{itemize}

\section{Appendix E}
\label{app51}
\textbf{Distribution of $\epsilon_t$, $\eta_t$}\\
$\epsilon$, $\eta_t$ are taken as two point random variables. Given two 
points and expectation, 
their variance is fixed. Similarly given two points and variance, their 
expectation is fixed. So, 
for a two point distribution expectation and variance cannot be arbitrarily 
chosen given two points. 
Here $\eta_t$ take values \{1, $\eta$\} and $\eta$ is chosen in such a way to 
match expectation and 
variance and so do for $E_t(\epsilon_t)$.
$\eta_t$ is simulated randomly from a distribution 
which satisfies the 
given expectation variance property. Since $V(\eta_t)$ is very low, 
$E_t(\eta_t)$ can be used as a 
random number simulator from the unknown distribution. 

\section*{Compliance with Ethical Standards}

Disclosure of potential conflicts of interest

Research involving Human Participants and/or Animals

Informed consent


\begin{thebibliography}{1}

\bibitem{allengale3}
Allen, F., Gale, D.: Understanding Financial Crises, OUP: Oxford, UK, 
Clarendon Lecture in Economics  (2007)

\bibitem{ban}
Banerjee, A. V.: A simple model of herd behaviour, Quarterly 
Journal of 
Economics 107, 797-817  (1992).

\bibitem{agh}
Aghion, P., Bacchetta, P., Banerjee, A.V.: A Simple Model of 
Monetary Policy and Currency Crises, European Economic Review, 44, 728-38  (2000)

\bibitem{gkb} 
Basak, G.K., Das, P.K., Marjit, S.:  Foreign capital inflow, 
exchange rate dynamics and potential financial crisis, Vol 103, 
page:673-684, Current science, India (2012)

\bibitem{Bellman} 
Bellman, R. and Dreyfus, S.:
Applied Dynamic Programming, Princeton University Press,
New Jersy {\em The Bellman principal of optimality, online site of Rosu, I.,
http://webhost.hec.fr/rosu/research/notes/bellman.pdf}  (1962).

\bibitem{bhag}
Bhagwati, J.: The Capital Myth: The Difference Between Trade in 
Widgets and Trade in Dollars, Foreign Affairs, 77, 7-12  (1998)

\bibitem{buss}
Bussi\`{e}re, M., Fratzscher, M., Koeniger, W.: Currency 
Mismatch, Uncertainty and Debt Maturity Structure, ECB Working Paper Series 
No. 409  (2004) 

\bibitem{calvo}
Calvo, G.A.: Capital Flows and Capital-Market Crises: The 
Simple Economics of 
Sudden Stops, Journal of Applied Economics, 1, 1, 35-54  (1998)

\bibitem{caves}
Caves, R. E., Frankel, J. A., Jones, R. W.: World Trade and 
Payments: An Introduction, Sixth Edition, Harper Collins College Publishers, 
New York (1993)

\bibitem{costinot}
Costinot, A., Lorenzoni, G., Werning, I.:
A Theory of Capital Controls as Dynamic Terms-of-Trade Manipulation,
Journal of Political Economy, 122, 1, 77-128  (2014)

\bibitem{dominq}
Dominguez, K. M. E., Tesar, L.L.: International Borrowing 
and Macroeconomic Performance in Argentina, NBER Working Paper No. 11353  (2005)

\bibitem{eichen1}
Eichengreen, B., Hausmann, R. : Exchange Rates and Financial
Fragility, Federal Reserve Bank of Kansas City, New Challenges for Monetary 
Policy, pp. 329-368 (1999)

\bibitem{eichen2}
Eichengreen, B., Hausmann, R.,  Panizza, U. : The Pain of
Original Sin, in Barry Eichengreen and Ricardo Hausmann (eds.), Other 
People's Money, University of Chicago Press: Chicago (2005)

\bibitem{gold}
Goldfajn, I.,  Minella, A.: Capital Flows and Controls in 
Brazil: What Have We Learned? NBER Working Paper No. 11640 (2005)

\bibitem{gorton}
Gorton, G.: The Panic of 2007, Federal Reserve Bank of Kansas City 
Symposium 2008 (available at 
http://www.kc.frb.org/publicat/sysmpos/2008/Gorton.10.04.08.pdf) (2008)

\bibitem{helpman}
Helpman, E., Krugman, P. A.: Trade Policy and Market Structure, MIT 
Press, Cambridge, Mass  (1989)

\bibitem{krugman}
Krugman, P. A.: What happened to Asia? Website of Paul A. Krugman (1998)

\bibitem{kindle}
Kindleberger, C.: Manias, Panics, and Crashes: A History of financial 
Crises, Basic Books: New York  (1978)

\bibitem{marjit}
Marjit, S., Das, P. K.,  Bardhan, S.: A portfolio based theory of 
foreign borrowing and capital control in a small open economy, Research in 
International Business and Finance, 21, 175-187  (2007)

\bibitem{noland}
Noland, M.: South Korea's Experience with International Capital 
Flows, NBER Working Paper No. 11381  (2005)

\bibitem{ostroy}
Ostroy, J. D., Ghosh,A. R., Habermeier, K.,
Chamon, M., Qureshi, M., Reinhardt, D. B.S.: Capital Inflows: The Role 
of Controls, IMF Staff Position Note No. SPN 10/04, IMF: Wasington D.C.  (2010) 

\bibitem{rakhit1}
Rakshit, M.K.: The East Asian Currency Crisis, OUP: New Delhi  (2002)

\bibitem{reinhart1}
Reinhart, V., Reinhart, C. M.: Capital Flows Bonanzas: AN 
Encompassing View of the Past and Present, CEPR Discussion 
Paper No. DP6996, UK  (2008)

\bibitem{reinhartrogoff}
Reinhart, C. M., Rogoff, K.: This Time is Different: Eight 
Centuries of Financial Folly, Princeton University Press, Princeton: New Jersy (2009)

\bibitem{rodrik}
Rodrik, D.: Who Needs Capital-Account Convertibility? Essays in 
International Finance 207, Princeton, New Jersey: Princeton University (1998)	

\bibitem{sen}
Sen, P.: Case Against Rushing into Full Capital Account Convertibility, Economic 
and Political Weekly, 41, 19, 1853-1857  (2006)

\bibitem{wolf}
Wolf, M.: Fixing Global Finance, John Hopkins Press, Baltimore (2008)

\bibitem{Uniroot} 
rootSolve package-http:// stat.ethz.ch / R-manual / R-patched / library / 
stats /html /
uniroot.html


\end{thebibliography}
\end{document}